\documentclass[twocolumn, twocolappendix]{aastex63}
\usepackage[italicdiff]{physics} 
\usepackage{tcolorbox}

\newcommand{\bs}[1]{\boldsymbol{#1}}

\usepackage[utf8]{inputenc}
\usepackage{amssymb,natbib,graphicx,color,ctable, amsmath,array,enumitem,kantlipsum}
\usepackage{hyperref}
\hypersetup{
    colorlinks=true,
    linkcolor=blue,
    citecolor=blue,
    filecolor=blue,
    urlcolor=blue
}

\usepackage[encapsulated]{CJK}
\usepackage{ucs}
\newcommand{\cntext}[1]{\begin{CJK}{UTF8}{bsmi}#1\end{CJK}}

\usepackage[flushleft]{threeparttable}
\usepackage{lineno}
\begin{document}

\defcitealias{HSvD21}{HSvD21}
\defcitealias{Ostriker2022}{OK22}

\title{The GHOSDT Simulations (Galaxy Hydrodynamical Simulations with Supernova-Driven Turbulence) - I. Magnetic Support in Gas Rich Disks}
\author[0009-0004-2434-8682]{Alon Gurman}
\affiliation{School of Physics \& Astronomy, Tel Aviv University, Ramat Aviv 69978, Israel}
\author[0000-0001-8867-5026]{Ulrich P. Steinwandel}
\affiliation{Center for Computational Astrophysics, Flatiron Institute, 162 5th Ave, New York, NY 10010, USA}
\author[0000-0002-9235-3529]{Chia-Yu Hu (\cntext{胡家瑜})}
\affiliation{Department of Astronomy, University of Florida, 211 Bryant Space Sciences Center, Gainesville, FL 32611 USA}
\affiliation
{Institute of Astrophysics and Department of Physics, National Taiwan University, No. 1, Sec. 4, Roosevelt Rd., Taipei 10617, Taiwan}
\author[0000-0001-5065-9530]{Amiel Sternberg}
\affiliation{School of Physics \& Astronomy, Tel Aviv University, Ramat Aviv 69978, Israel}
\affiliation{Center for Computational Astrophysics, Flatiron Institute, 162 5th Ave, New York, NY 10010, USA}
\affiliation{Max-Planck-Institut f\"{u}r Extraterrestrische Physik, Giessenbachstrasse 1, D-85748 Garching, Germany}
\correspondingauthor{Alon Gurman}
\email{alongurman@gmail.com}

\begin{abstract}
Galaxies at redshift $z\sim 1-2$ display high star formation rates (SFRs) with elevated cold gas fractions and column densities. Simulating a self-regulated ISM in a hydrodynamical, self-consistent context, has proven challenging due to strong outflows triggered by supernova (SN) feedback. At sufficiently high gas column densities, if magnetic fields or other mitigating measures are not implemented, these outflows can prevent a quasi-steady disk from forming for several 100 Myr. To this end, we present 
GHOSDT, 
a suite of magneto-hydrodynamical simulations that implement ISM physics at high resolution. We demonstrate that magnetic pressure is important in the dense ISM of gas-rich star-forming disks. We show that a relation between the magnetic field and gas surface density emerges naturally from our simulations. We argue that the magnetic field in the dense, star-forming gas, may be set by the SN-driven turbulent gas motions.
When compared to pure hydrodynamical runs, we find that the inclusion of magnetic fields increases the cold gas fraction by up to 40\%, reduces the disc scale height by up to a factor of $\sim 2$, and reduces the star formation burstiness. In dense ($n>100\;\rm{cm}^{-3}$) gas, we find steady-state magnetic field strengths of 10--40 $\mu$G, comparable to those observed in Galactic molecular clouds. Finally, we demonstrate that our simulation framework is consistent with the \citet{Ostriker2022} Pressure Regulated Feedback Modulated Theory of star formation and stellar feedback.
\end{abstract}


\section{Introduction}

Stars form in cold, dense, molecular clouds \citep{McKee2007,Leroy2008,Tacconi2020,Chevance2023}. Static models of the multiphase interstellar medium (ISM) capture the relationship between the ISM and star formation, in which the phase distribution and gas conditions of the ISM are affected by energy injection from stars in the form of radiative- and supernova feedback \citep{Klessen2016, Naab2017, Bialy2019}. The time-dependent picture, however, is more complex, as dynamical processes such as the cycle of cooling, gravitational collapse, and subsequent radiative feedback, shock heating, and dispersal of clouds, take place rapidly and are affected by the large-scale gravitational potential, gas weight, and star-formation-rate (SFR), to name a few \citep[e.g.,][]{Ostriker2011,HSvD21,Ostriker2022}. As such, dynamical models of the ISM in the form of 3D (magneto-)hydrodynamical simulations on different scales have been devised and implemented at high resolution on different physical scales. Stratified box setups, also referred to as galactic patches, strike a balance between simulation volume, spatial resolution, and simulation time. They combine (sub-)pc resolution with a kpc-scale box size, and several 100 Myr of simulation time, and provide several benefits \citep[see, e.g.,][]{DeAvillez2004b, Joung2006, Hill2012, Hennebelle2014, Walch2015b, Girichidis2015a, Kim2015, Simpson2016, Hu2016, Hu2017, Kim2018, Kim2020, Kim2020b, Kannan2020, HSvD21, Brucy2023, Rathjen2023, Kim2024, Rathjen2024}. This treatment allows for the (potential) resolution of the dense, clumpy, molecular cores of star-forming clouds, with a large enough volume for capturing gas compression on the scale of several 100 pc by SN shocks. The long simulation time means that several life cycles of molecular clouds can be captured, and stochastic nature properly sampled. In addition, it allows averaging large fluctuations of different ISM properties over time. Finally, at this scale, star formation can be tracked on a star-by-star level. While this requires additional assumptions, such as the adoption of an initial mass function (IMF), stars can be formed individually, and their associated feedback can be treated without the need for stellar population synthesis. This approach was first implemented in \citet{Hu2016} in simulations of dwarf galaxies. A second widely adopted approach in high-resolution simulations of galactic patches is to form star clusters in the form of sink particles rather than single stars \citep[e.g.,][]{Kim2017, Walch2015b}. Following their formation, sink particles are allowed to accrete gas from their environment. SN and photoionization feedback are then computed based on the accretion history of the sink particles.

\citet[][hereafter \citetalias{HSvD21}]{HSvD21} performed a suite of hydrodynamical simulations of a $\left(1\times1\times20\right)\;\mathrm{kpc}^3$ ISM patch. They used a mass resolution of 1 $M_{\odot}$, and a Jeans-mass based criterion for star formation. This leads to a characteristic density of $10^4\;\rm cm^{-3}$ at which star formation takes place. At this density, the radius of the meshless finite-mass (MFM) kernel is $\sim0.2$ pc, corresponding to the effective spatial resolution at this density. They ran their simulations for 500 Myr, and 4 different metallicities ranging from $Z/Z_{\odot}=0.1$ to 3, and including a star-by-star treatment of star formation and SN feedback, their setup was optimal for investigating the properties of the cold (molecular) ISM. Comparing a steady-state and a time-dependent treatment of hydrogen chemistry, they found that assuming chemical steady state overproduces H$_2$, and that this overproduction effect is stronger at low metallicity, where the H$_2$ formation timescale is longer. By using full line radiative transfer, \citet{Hu2022a} found that this effect leads to a substantial correction to the widely assumed metallicity dependence of the CO-to-H$_2$ conversion factor, $X_{\mathrm{CO}}$. Based on the \citet{Hu2022a} simulations,  \citet{Gurman2024} found a linear metallicity dependence in the [$\text{C\,\small{II}}$]--SFR relation, with [$\text{C\,\small{II}}$] emission dominated by high-density gas, consistent with observations \citep{Pineda2014,Croxall2017}.

On average, galaxies at $z\sim 1-2$, form stars at higher rates per stellar mass compared to those at $z \sim 0$ \citep{Madau2014}, with Milky Way mass galaxies peaking near $z\sim1$ \citep{Noeske2007,Whitaker2014,Moster2018, ForsterSchreiber2020}. The cosmic peak of star-formation at $z\sim 1-2$ is associated with larger cold gas fractions and gas column densities compared to $z=0$ \citep{Tacconi2020}. 

The galactic patch simulations predict strong SN-driven gas outflows at gas surface densities higher than the solar neighborhood value of the $\sim$~10~M$_\odot$~pc$^{-2}$. The increase in outflows is in direct correlation to the gas surface density, which, in turn, causes an increase in SFR as is observed in the Kennicutt-Schmidt relation on both ISM and galaxy-integrated scales \citep{Kennicutt1998b,Bigiel2008}. In simulations, these strong outflows can potentially disrupt the emerging gaseous disks entirely, making them challenging to simulate. Successful simulations at this scale with high gas-surface densities have been presented by \citet{Kim2020, Rathjen2023, Brucy2023}. In all of the above, a combination of magnetic fields and (temporary) artificial turbulent driving have been implemented to moderate the large-scale, simultaneous, vertical collapse of the disk, and the subsequent intense SN-driven outflows. This enables the retention of much of the gas in the box for a longer period.
In the absence of the aforementioned precautions, as in the setup of \citetalias{HSvD21}, these high gas surface density simulations are known to blow out most of their gas, while the remaining gas fails to settle into a star-forming disk, and the star formation rate falls well below the observed Kennicutt-Schmidt relation. It is in this context that we wish to explore the roles of magnetic fields in stabilizing a stratified ISM disk against collapse, regulating star formation, and determining the phase distribution of the gas.

The effects of magnetic fields in a self-regulated ISM have been studied at solar neighborhood conditions, i.e., $\Sigma_{\rm{gas}}\approx 10 \; M_{\odot} \; \rm{pc}^{-2}$. \citet{Hill2012} compared the ISM phase structure in pure hydrodynamical simulations to that of magnetohydrodynamical (MHD) simulations with different initial magnetic fields. They found that the cold gas mass fractions are lower for the MHD runs, likely due to magnetic support against the vertical compressions of their initial disk setups. However, they did not include self-gravity, temporal and spatial variations in the FUV radiation field, or a self-consistent positioning of the SN explosions. \citet{Hennebelle2014} include self-gravity in their simulations, and position an SN explosion in the vicinity of stellar sink particles whenever they accrete  120 $M_{\odot}$. While their simulation time of $\lesssim 100$ Myr is rather short, they find that their hydrodynamical run forms stars in shorter, more intense bursts, compared to the MHD runs, while the total mass in stars formed is reduced slightly by the inclusion of magnetic fields, provided the initial field is strong enough. Using a similar setup \citet{Iffrig2017} explored higher initial magnetic field strengths. They found a reduction of up to an order of magnitude in the SFR for the MHD simulations compared to pure hydrodynamical simulations. The simulation time of $\lesssim 100$ Myr limited their study to the initial burst of star formation following the vertical collapse of their initial disk setup. They also found that the orientation of the magnetic field, initially set to be uniform and in the disk plane, mostly retained its initial direction even at the end of their simulations, indicating the persistence of the initial magnetic field configuration. In a comparison between hydrodynamical and MHD ISM patches with different initial magnetic field strengths and configurations, \citet{Kim2015} found that stronger magnetic fields reduce the scale height of the disk as well as the SFR. They note that in their strongest magnetic field runs, the magnetic field strength does not reach steady state, and was still dominated by the value set by the initial conditions, even at late times. \citet{Girichidis2018} found that including magnetic fields leads to a larger disk scale height, but this is likely due to their simulation time ($\sim 60$ Myr) being limited to the collapse of their initial disk setup. As such, magnetic support simply slows down the vertical collapse and creates an effectively thicker disk. This also explains why their MHD runs show a delayed onset of H$_2$ and star formation, as both are correlated with the compression of the disk. They also found that the relationship between gas density and magnetic field strength is almost unaffected by a factor of two change in the initial magnetic field. 
\cite{Gressel2008} simulated a SN-regulated ISM in a shearing box with resistive MHD and a spatially constant magnetic diffusivity. Their fiducial setup included Coriolis forces to model the effect of differential rotation, and they also ran simulations with Coriolis forces disabled. \cite{Gressel2008} found that while both displayed an amplification of the root mean square magnetic field, the mean magnetic field could only be amplified when differential rotation was included in the form of Coriolis forces.
\citet{Brucy2023b} find that for high gas surface density, increasing the magnetic field can reduce the SFR by 2-3 orders of magnitude. It should be noted that due to the bursty star formation in their gas-rich models, they restricted their analysis in these cases to the first, strong burst of star formation. Since this is the first event of star formation, the gas has yet to experience regulation by SNe, and thus the SFR could become rather sensitive to initial conditions.

Some studies of the ISM focus on the evolution of magnetic fields and in particular the effect of small-scale turbulent dynamo (SSD) using different setups. \cite{Gent2021,Gent2023} conducted simulations of SN-driven periodic ISM volume, finding that the overall magnetic energy saturated at a value of 5\% of the kinetic energy, while in the cold gas ($T<3000\ \rm{K}$) the magnetic energy was in equipartition with the kinetic energy. \cite{Rieder2017} simulated an isolated dwarf galaxy at high-resolution, demonstrating a SN-driven SSD saturating at 3\% of the total kinetic energy. \cite{Pakmor2017} presented zoom simulations of Milky Way mass halos, where the initial and boundary conditions are set by a lower resolution, cosmological simulation. They found that at saturation, the magnetic energy was equal to 20\% of the kinetic energy in the central 1 kpc of the halos, and dropped to 5-10\% at distances of 10-30 kpc.

To explore the effects of magnetic fields on the ISM over a range of gas surface densities, we present
GHOSDT (Galaxy Hydrodynamical Simulations with Supernova-Driven Turbulence), 
a suite of high-resolution, magneto-hydrodynamical ISM simulations, spanning gas surface densities in the range of 4-100 $M_{\odot}\;\rm{pc}^{-2}$. In Section~\ref{section: methods}, we describe our numerical setup and different simulation runs. In Section~\ref{section: results}, we present an overview of the simulation results and the effect of magnetic fields. In Section~\ref{section: B fields tests}, we investigate the dependence of the magnetic field on the gas surface density and initial conditions. In Section~\ref{sec: summary}, we summarize our results and discuss future  prospects for the simulations.

\section{Numerical Methods}
\label{section: methods}
GHOSDT
builds on the simulations of \citetalias{HSvD21}, and follows the same implementations for cooling, heating, chemistry, star formation, and stellar feedback. We extend the \citetalias{HSvD21} setup by including magnetic fields and varying the initial gas surface density, to study the higher column density regime.

\subsection{Gravity and MHD}
We employ the publicly available version of \textsc {Gizmo} \citep{Hopkins2015}, a multi-method code implementing the meshless Godunov-type method \citep{Gaburov2011} on top of the TreeSPH code \textsc{Gadget}-3 \citep{Springel2005}. The gravitational interaction is solved by the “treecode” method outlined in \citet{Barnes1986}, while hydrodynamics are solved by the MFM method \citep{Hopkins2015} with the (average) number of neighboring particles in a kernel $N_\mathrm{ngb} = 32$. 

We extend the work of \citetalias{HSvD21} by also including magnetic fields using the \textsc{Gizmo} module introduced by \citet{Hopkins2016}, solving the (ideal) MHD equations, while ensuring numerical stability and maintaining $\nabla \cdot \bs{B} \approx 0$ (see Appendix C). This is done using a combination of the Powell 8-wave cleaning \citep{Powell1999} and the hyperbolic/parabolic divergence cleaning scheme of \citet{Dedner2002} that we briefly introduce below. While we do not employ the constrained-gradient (CG) approach presented in \citet{Hopkins2016b}, in Appendix C we confirm their findings that this method reduces the error in $\grad \cdot \bs{B}$ by $\sim1$ dex, providing motivation for implementation of CG in future studies.

The homogeneous ideal MHD Euler equations for a moving frame with velocity $v_{\rm{frame}}$ are given by
\begin{equation}
    \frac{\partial \bs{U}}{\partial t} + \grad \cdot \left( \bs{F} - \bs{v}_{\rm{frame}} \otimes \bs{U} \right)=\bs{S}
\end{equation}
where $\bs{U}$ is the vector of conserved variables (in the source-free case), $\bs{F}$ is the tensor consisting of the different fluxes of each conserved variable, and $\bs{S}$ is the source term vector.

\begin{equation}
    \bs{U}=\begin{pmatrix}
    \rho \\ \rho\;\bs{v} \\ \rho \;e \\ \bs{B} \\ \rho\; \psi 
    \end{pmatrix}
    \\;\;\;
    \bs{F}=\begin{pmatrix}
        \rho \;\bs{v} \\ \rho\; \bs{v} \otimes v + P_T \mathcal{I} - \bs{B} \otimes \bs{B} \\ \left(\rho \;e + P_T \right) \bs{v} - \left(\bs{v} \cdot \bs{B} \right) \bs{B} \\ \bs{v}\otimes \bs{B} - \bs{B}\otimes \bs{v} \\ \rho \; \psi \; \bs{v}
    \end{pmatrix}
\end{equation}
where $\rho$ is the mass density, $e=u+\left| \bs{B} \right| ^2/2\rho + \left|\bs{v}\right|^2/2$ is the total specific energy, $u$ is the internal energy, $P_T=P+\left|\bs{B}\right|^2/2$ is the sum of the thermal and magnetic pressures, and $\psi$ is a conservative scalar field devised to transport $\grad \cdot \bs{B}$ away from the source and to damp it, following the method of \citet{Dedner2002}.

The discrete meshless evolution equations, fully derived in \citet{Hopkins2015}, are then given by
\begin{equation}
    \frac{d}{dt} \left( V\;\bs{U} \right) _i + \sum_j \; \tilde{\bs{F}}_{i,j} \cdot \bs{A}_{ij}=\left(V\;\bs{S}\right)_i.
\end{equation}
Here, $\left( V\;\bs{U}\right)_i$ is the cell-volume integrated valued of conserved quantities to be carried with particle $i$, and its rate of change is given by the sum of fluxes $\tilde{\bs{F}}$ into/out of its effective face area $\bs{A}_{ij}$ and the cell-integrated source term $\left(V\;\bs{S}\right)_i$. The full derivation of $\bs{A}_{ij}$ is presented in \citet{Hopkins2015}.

The cleaning method described in \citet{Hopkins2016} adds two source terms,
\begin{equation}
    \bs{S}=\bs{S}_{\rm{Powell}}+\bs{S}_{\rm{Dedner}},
\end{equation}
\begin{equation*}
    =-\grad \cdot \bs{B} \begin{pmatrix}
    0 \\ 
    \bs{B} \\ 
    \bs{v}\cdot \bs{B} \\ 
    \bs{v} \\ 
    0

    \end{pmatrix} - \begin{pmatrix}
    0 \\ 
    0 \\ 
    \bs{B}\cdot \left( \grad \psi \right) \\ 
    \grad \psi \\ 
    \left(\grad \cdot \bs{B} \right) \rho \;c_{\rm{h}}^2+\rho \;\psi / \tau

    \end{pmatrix}.
\end{equation*}
$\bs{S}_{\rm{Powell}}$ represents the \citet{Powell1999} cleaning, which subtracts the numerically unstable terms resulting from a non-zero $\nabla \cdot \bs{B}$. $\bs{S}_{\rm{Dedner}}$ is the source term formulated in \citet{Dedner2002}, where a scalar field $\psi$ is introduced in order to transport away and damp the non-zero divergence. There is freedom to choose $c_{\rm{h}}$ and $\tau$ for each particle. For each particle, we use $c_{\rm{h}}=\sigma_{\rm{h}}^{1/2} v_{\rm{sig}}^{\rm{MAX}}/2$ (following \citet{Hopkins2016, Tricco2012}), where $v_{\rm{sig}}^{\rm{MAX}}$ is the local maximum signal velocity, and $\sigma_{\rm{h}}$ is a dimensionless scaling parameter taken to be 1. We use $\tau=h/\left( \sigma_{\rm{p}}c_{\tau} \right)$, where $h$ is the local effective cell height, $\sigma_{\rm{p}}=0.1$, and the damping speed $c_{\tau}$ is defined in Appendix D of \citet{Hopkins2016}, where the choices of definitions for the above quantities is explored in depth.

\subsection{ISM Physics and Star Formation}
Our treatment of cooling, chemistry, star formation, and feedback, uses the setup presented in \citetalias{HSvD21}. Chemistry and cooling are based on \citet{Glover2007} and \citet{Glover2012} and include time-dependent treatment of hydrogen chemistry. 
This is combined with a HEALPIX \citep{Gorski2011} method for the calculation of radiation shielding, which captures H$_2$ and dust FUV shielding by determining an effective shielding column density for each particle.

Star formation is implemented using a stochastic recipe. The conditions for star formation are that the local velocity divergence becomes negative (i.e., converging flows) and the local thermal jeans mass becomes unresolved, i.e., lower than the local kernel mass. When these conditions are met, a gas particle is assigned a probability of $\epsilon_{\mathrm{sf}}\Delta t / t_{\mathrm{ff}}$ to be converted into a star particle within a single time step, where $\Delta t$ is the time step, $t_{\mathrm{ff}}$ is the local freefall time, and $\epsilon_{\mathrm{sf}}$ is the star formation efficiency, assumed to be equal to 0.5, as we resolve the clumpy sub-structure of molecular clouds. Additionally, if the density of a gas particle exceeds a threshold density, it is converted into a star particle instantaneously, irrespective of the previously listed conditions. Our adopted threshold density for instantaneous star formation is $n_{\rm{isf}}=10^5\;\rm{cm}^{-3}$. The mass of the star is sampled from a \citet{Kroupa2002} IMF, and this determines its photoionizing budget and whether it produces SN feedback. In the gravity calculations that follow, the newly formed star particle maintains the mass of its parent gas particle. This apparent mismatch between gas mass converted into stars and the stellar mass drawn from the IMF disappears when averaged over a large number of stars, ensuring that the ratio between the SN rate and the SFR is appropriate for the chosen IMF.

If a star has mass $>8\ M_{\odot}$, we use the method presented in \citet{Hu2017}, where a spherical ionization front within which hydrogen is assumed to be fully ionized, and the gas is heated to $10^4$ K, is computed iteratively until the (initial-mass dependent) ionizing photon production rate of the star (or stars) equals the total recombination rate in the ionized region. Stars with mass $>8\ M_{\odot}$ also inject $10^{51}$ ergs of thermal energy into their neighbouring 100 particles at the end of their lifetime, which is determined by their initial mass as well. \citet{Hu2019b} demonstrated that when the Sedov-Taylor phase of a SN remnant expansion is resolved, the SN feedback recipe is insensitive to whether we choose a thermal energy or momentum injection scheme. In the momentum injection scheme, neighbouring particles are assigned a density-dependent momentum, which is calibrated against the terminal momentum of the momentum-conserving phase in idealized simulations of an isolated SN remnant expansion. \citet{Steinwandel2020} showed that that the terminal momentum of the SNR is accurately captured with a purely thermal energy injection scheme (as is used in this work) with a mass resolution of $10\;M_{\odot}$, while the wind thermal properties require better resolution regardless of whether or not a momentum injection scheme is applied. Thus, we do not apply any momentum injection in our SN feedback model.

This star-by-star approach to high-resolution ISM simulation introduced in \citet{Hu2016} was subsequently implemented in the codes 
\textsc{Gadget} \citep{Hu2017} and \textsc{Gizmo} \citep{Hu2021}. It has since been adopted as the default star formation scheme in additional models, e.g., \citet{Hu2019b}, \citet{Lahen2020}, \citet{Steinwandel2020}, \citet{Hu2023b}, \citet{Partmann2024}. It has also been implemented in other multi-purpose hydrodynamical codes such as \textsc{Enzo} \citep{emerick2019}, \textsc{Arepo} \citep{Smith2021}, \textsc{Ramses} \citep{Andersson2020}, and \textsc{Asura+Bridge} \citep[][]{Hirai2021}.

We do not account for the evolution of stellar winds, which may impact the structure and extent of photoionized regions before SN feedback disrupts them entirely \citep{Haid2018, Lancaster2021, Lancaster2021b, Grudic2021, Grudic2022}.

\subsection{Simulation Setup}
\label{section: simulation setup}

Our simulation setup consists of a spatial domain extending 1 kpc in the planar $x$- and $y$-directions, and 50 kpc in the vertical $z$-direction. Boundary conditions are periodic in $x$ and $y$, and the $z$-dimension has outflow boundary conditions. The mid-plane of the disk is set at $z=0$. In our fiducial initial conditions, the gas initially follows a \citet{SpitzerLyman1942} vertical distribution of the form
\begin{equation}
\label{eq:spitzer}
    \rho\left(z\right)=\Sigma_{\rm{gas,0}}/\left(2H_g\right)\mathrm{sech}^2 \left( z/H_g \right) \ \ \ .
\end{equation}
Here $\Sigma_{\rm{gas,0}}$, the initial total gas surface density, is the main free parameter in this work. We consider the range 4 to 100 $M_{\odot}\;\rm{pc}^{-2}$. In Eq.~(\ref{eq:spitzer}), $\rho$ is the gas density, and $H_g$ is the initial scale height of the disk which we set to 250 pc. At any height $z$, the initial acceleration due to self-gravity is
\begin{equation}
    a_g=-2\pi \ G \ \Sigma_{\mathrm{gas,0\;}}\mathrm{tanh}\left(\frac{z}{H_g}\right) \ \ \ .
\end{equation}
We also include an external gravitational potential due to a stellar disk and dark matter halo. The contribution to the gravitational acceleration from the stellar disk is
\begin{equation}
    a_*=-2\pi G \Sigma_* \mathrm{tanh}\left( \frac{z}{H_*} \right) \ \ \ ,
\end{equation}
where $\Sigma_*=40\;M_{\odot}\;\mathrm{pc}^{-2}$ and $H_*=250$ pc are typical values at the Galactic solar circle. The dark matter contribution is derived from an Navarro–Frenk–White profile \citep[NFW;][]{Navarro1997} with a virial mass $M_{\mathrm{vir}}=10^{12}\;M_{\odot}$ and concentration $c=12$, resulting in
\begin{equation}
    a_{\mathrm{DM}}=-\frac{Gm\left(r\right)z}{r^3},
\end{equation}
where $r=\sqrt{z^2+R_0^2}$ is the radial distance from the center of the halo. The enclosed mass $m\left(r\right)$ is given by
\begin{equation}
    m\left(r\right)=4\pi r_s^3 \rho_s \ln \left[\left(1+r/r_s\right)-\left(r/r_s\right)\left(1+r/r_s\right) \right],
\end{equation}
where $r_s=17$ kpc, $\rho_s=9.5\times 10^{-3}\;M_{\odot}\;\mathrm{pc}^{-3}$, and $R_0=8$ kpc is the galactocentric distance to the Sun. We keep both $a_*$ and $a_{\rm{DM}}$ fixed across all of our simulations, independent of the assumed gas surface density. As the external potential is dominated by the stellar component, we can estimate the vertical oscillation time associated with it at a vertical distance $z$ from the midplane, and assuming $z\gg H_{*}$

\begin{equation}
    t_{\rm{ver,*}}\approx 4 \times \sqrt{\frac{2z}{a_*}}=\left( \frac{z}{1\;\rm{kpc}} \right)^{0.5}\times 168 \;\rm{Myr} \ \ \ .
\end{equation}

Our mass resolution is $m_{\rm g}=1\;M_{\odot}$ for our high-resolution models and $m_{\rm g}=10\;M_{\odot}$ for our low-resolution models, with the exception of $\Sigma_{\rm{gas,0}}=60\;M_{\odot}\;\rm{pc}^{-2}$. At this gas surface density, our high-resolution model was run with $m_{\rm g}=4\;M_{\odot}$, due to computation time considerations, and we group it with our 1 $M_{\odot}$ resolution models in most of our analysis. At this resolution, the Sedov-Taylor phase of a SN remnant expansion is captured accurately \citep{Steinwandel2020}. While high resolution is desirable for the correct treatment of SN feedback and resolving the structure of molecular clouds, the computational demands of such high resolution become very high for $\Sigma_{\rm{gas,0}}>40\;M_{\odot}\ \rm{pc}^{-2}$, and we are forced to use lower resolution for those models, setting the gas particle mass to 10 $M_{\odot}$. For consistency, we run low-resolution models for $\Sigma_{\rm{gas,0}}\leq 40\;M_{\odot}\ \mathrm{pc}^{-2}$ as well. At the density corresponding to our star-formation threshold, this corresponds to a spatial resolution of $\approx 0.2 $ pc for our $1\;M_{\odot}$ runs, or $\approx 2$ pc for our $10\;M_{\odot}$ runs. As is the case for Lagrangian codes, our resolution is degraded at lower density gas and scales as $n^{-1/3}$, where $n$ is the gas volume density. We run our models for 0.5-1 Gyr. Our long simulation times allow us to fully sample the fluctuations in star formation and ISM conditions, as well as the evolution of gas mass and magnetic fields over time. 

Only a small number of groups have carried out galactic patch simulations at high resolution with a multi-channel stellar feedback with MHD and high gas surface density models that are evolved past the initial burst of star formation that is common in these systems \citep[e.g.,][]{Kim2020}. Also, our simulations include pure hydrodynamical models up to $\Sigma_{\rm{gas,0}}=80\;M_{\odot}\;\rm{pc}^{-2}$, allowing us to investigate the importance of magnetic fields in these environments.

The initial gas temperature is set to $10^4$ K everywhere. We adopt a constant solar chemical composition, i.e., carbon and oxygen abundance of $1.4\times 10^{-4}$ and $3.2\times 10^{-4}$, respectively \citep{Cardelli1996, Sembach2000}, and a dust-to-gas mass ratio of 1\%. 

In our fiducial MHD runs, the magnetic field is set to an initial value $B_0 \hat{x}$ (parallel to the disk mid-plane), which is constant everywhere. We choose the value of $B_0$ by attempting to find a value that fulfills the following requirements. First, it should allow for the establishment of a quasi-steady disk, which becomes increasingly important as $\Sigma_{\rm{gas},0}$ increases (see Section~\ref{sec: star formation}). Second, we require that the mass-weighted mean magnetic field in the dense ($n>100\;\rm{cm}^{-3}$) gas reaches a quasi-steady value. Since the magnetic field strength varies in time following fluctuations in SFR, we first apply a moving average with a 200 Myr window, and then visually check for convergence. Third, we require that the absolute value of the $x$- and $y$-components of the magnetic become comparable. We do so by requiring that the mass-weighted average of $\left( \bs{B}_x/\left|\bs{B}\right| \right)^2$ decreases to a value close to 1/3, roughly indicating equipartition of the different magnetic field components. We show that this holds for our simulations in Section~\ref{section: B fields tests} and Appendix B. Finally, we check that the mean magnetic field is negligible compared to the root mean square magnetic field. Under these conditions, the simulation results are quite insensitive to the value of $B_0$, even when it is varied by orders of magnitude. We briefly discuss the limitations and potential for improvement in our choice of initial conditions in Appendix A.

\subsection{Simulation List}
\label{section: sim list}
Our simulation runs are labeled and listed in Table~\ref{table: simulation list}. Simulation groups H and M use our stratified disk initial conditions described in Section~\ref{section: simulation setup} with different values of $\Sigma_{\mathrm{gas,0}}$. Simulations in group H are run with magnetic fields switched off, while simulations in group M include magnetic fields whose initial $x$-component is set to $B_0$. Simulations in groups HV and MV are run with magnetic fields turned off and on, respectively, but with initial conditions that differ from groups H and M. Instead of a stratified disk, the initial conditions for groups with added label V are set using the outputs from the simulations of group M at 200 Myr, with the corresponding value of $\Sigma_{\rm{gas},0}$. Then, in group HV, the simulation is restarted with magnetic fields turned off. In group MV, each directional component of the initial magnetic field at each particle is set by drawing from a Gaussian random distribution with a standard deviation equal to $B_0$. While these initial conditions are not designed to satisfy the condition $\nabla \cdot \bs{B}=0$, the employed divergence cleaning scheme ensures that the error in $\nabla \cdot \bs{B}$ becomes comparable to that of our fiducial runs within $\lesssim$1 Myr. We demonstrate this in Appendix C. The aim of groups HV and MV is to use initial conditions that are more turbulent and less symmetrical, compared with the setup in groups H and M. 

When referring to a single simulation, we specify the value of $\Sigma_{\rm{gas},0}$ in units of $M_{\odot}\;\rm{pc}^{-2}$ following the group name (e.g., M20 belongs to group M and has $\Sigma_{\rm{gas},0}$ set to $20\;M_{\odot}\;\rm{pc}^{-2}$). In addition, we supplement the simulations in Table~\ref{table: simulation list} with an additional set of simulations whose purpose is to verify the insensitivity of the emerging magnetic field to our initial conditions. These simulations are listed and discussed in Section~\ref{section: B fields tests}. Unless explicitly stated otherwise, we only analyze our simulation data at times later than 200 Myr, to mitigate the effects of our initial conditions. This timescale is a combination of the free-fall and cooling times of the initial disk, followed by strong outflows, and then a second round of inflows, after which the disk can emerge. This timescale also changes for different simulations, with the low $\Sigma_{\rm{gas},0}$ runs generally taking longer for the initial collapse. 200 Myr is generally long enough to capture this period across all of our different simulations. Simulations that were run for a much shorter time period, like some existing works in the literature, are significantly affected by the initial conditions.  

\begin{table*}

\centering
\begin{threeparttable}

\centering 
\begin{tabular}{l l l l l}
\hline \hline 
\\ [-1.5ex] Name & $\Sigma_{\rm{gas,0}}$ ($M_{\odot}\;\rm{pc}^{-2}$) & $B_0$ ($\mu \rm{G}$)& $m_{\mathrm{g}}$ ($M_{\odot}$) & $t_{\mathrm{sim}}$ (Gyr)\\ [0.5ex] 
\hline
\\[-3ex]H4, H10, H20 & 4, 10, 20 & N/A & 1 (10) & 0.5 (1) 
\\[0.5ex] H40, H60, H80& 40, 60, 80 & N/A & 10 & 0.5 \\[0.5ex]
\hline
\\[-3ex]M4 & 4 & 0.05 & 1 (10) & 1
\\[0.5ex]M10 & 10 & 0.5 & 1 (10) & 0.5 (1)
\\[0.5ex]M20 & 20 & 2 & 1 (10) & 0.5 (1)
\\[0.5ex]M40 & 40 & 10 & 1 (10) & 0.5 (1)
\\[0.5ex]M60 & 60 & 20 & 4 (10) & 0.5 (1)
\\[0.5ex]M80 & 80 & 40 & 10 & 0.5
\\[0.5ex]M100 & 100 & 40 & 10 & 0.8
\\[0.5ex]
\hline
\\[-3ex]HV40, HV60, HV80 & 40, 60, 80 & N/A & 10 & 0.5 \\[0.5ex]
\hline
\\[-3ex]MV4 & 4 & 0.05, $5\times 10^{-4}$ & 10 & 0.5
\\[0.5ex]MV60 & 60 & 0.05 & 10 & 0.5 \\[0.5ex]
\hline

\end{tabular}

\end{threeparttable}
\caption{Overview of simulation parameters.}
\label{table: simulation list}
\end{table*}

\section{Results}
\label{section: results}

\subsection{Star Formation Rate}

\label{sec: star formation}

\begin{figure}	
	\centering
	\centerline{\includegraphics[trim={0 0 0 0},clip,width=0.8\linewidth]{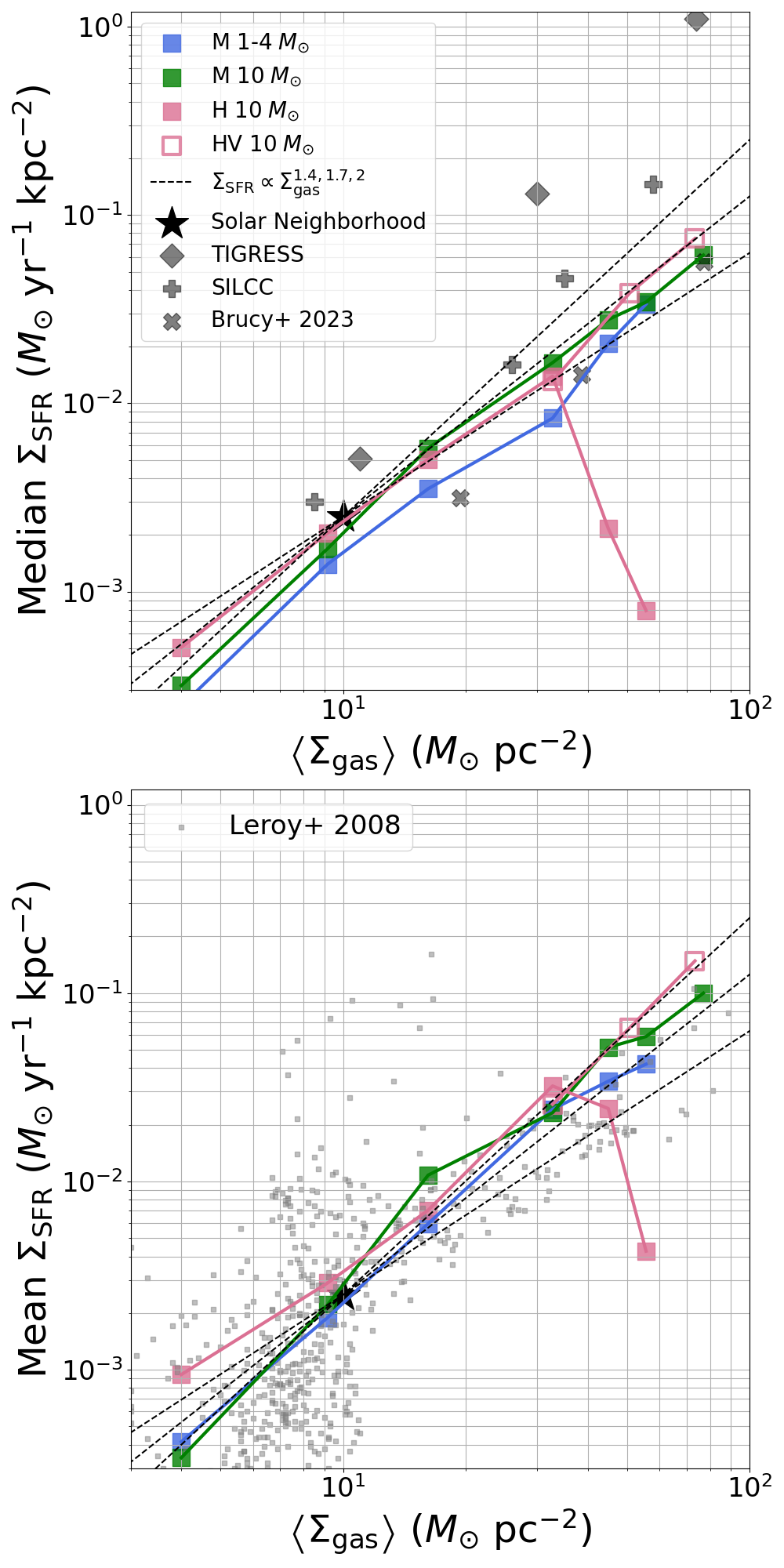}}
	\caption{
Median (top panel) and mean (bottom panel) binned values of $\Sigma_{\rm{SFR}}$ as a function of $\left<\Sigma_{\rm{gas}}\right>$ for groups H, M, and HV. Also plotted are observations from \citet{Leroy2008} and theoretical results from TIGRESS \citep{Kim2020}, SILCC \citep{Rathjen2023}, and \citet{Brucy2023}.
		}
		\label{fig: KS overview}
\end{figure}

Fig.~\ref{fig: KS overview} shows the average Kennicutt-Schmidt (KS) relation between the SFR surface density $\Sigma_{\rm SFR}$ ($M_\odot$ yr$^{-1}$ kpc$^{-2}$) and the gas surface density $\Sigma_{\rm gas}$ ($M_\odot$ pc$^{-2}$), for our simulation groups H, M, and HV. For each snapshot, we set
\begin{equation}
    {\rm SFR}=\frac{M_*(<t_{\rm SFR})}{t_{\rm SFR}} \ \ \ ,
\end{equation}
where $M_*(<t_{\rm SFR})$ is the mass in stars with ages less than an adopted star-formation timescale $t_{\mathrm{SFR}}=10$ Myr. We further divide the SFR by our box surface area of 1 kpc$^2$ to obtain  $\Sigma_{\rm{SFR}}$. Because the gas surface density decreases with time as gas is driven out of the box boundaries by SN feedback, especially for $\Sigma_{\rm{gas},0}\geq 40\;M_{\odot}\;\rm{pc}^{-2}$, for each group we bin individual snapshots taken at time intervals of 1 Myr according to their gas surface density.
Next, we compute the medians and means of $\Sigma_{\rm{SFR}}$ for the snapshots in each bin. Results are presented in the top and bottom panels of Fig.~\ref{fig: KS overview}, respectively. We also plot power laws of the form $\Sigma_{\rm{SFR}}\propto\Sigma_{\rm{gas}}^{N}$ for $N=(1.4,1.7,2)$, the observed solar neighborhood value from \citet{Fuchs2009}, and results from the TIGRESS simulation R-models \citep[][]{Kim2020}, group TURB from \citet{Brucy2023}, and the SILCC project \citep{Rathjen2023}.

\begin{figure*}	
	\centering
	\centerline{\includegraphics[trim={0 0.1cm 0 0},clip,width=1.3\linewidth]{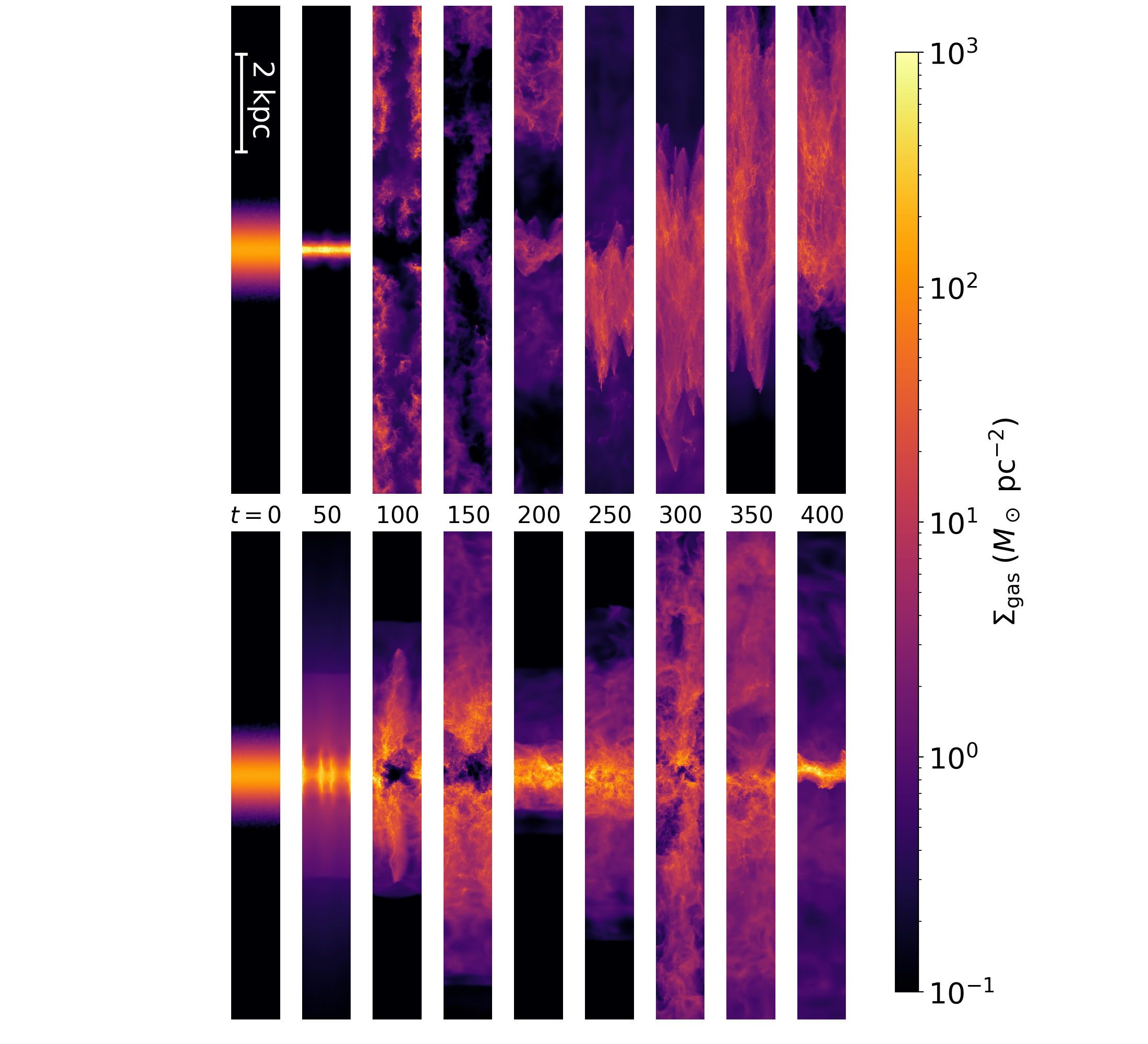}}
	\caption{
Edge-on gas surface density maps for snapshots at different times for simulations H80 (top row) and M80 (bottom row). Time is given in Myr.
		}
		\label{fig: visual comparison}
\end{figure*}

We find that our different simulation groups agree in their KS relations to within a factor of $\sim2$, except group H which shows a decrease in $\Sigma_{\mathrm{SFR}}$ for $\Sigma{\mathrm{gas}}\gtrsim 40 \; M_{\odot}\;\mathrm{pc}^{-2}$. These small differences do not seem systematic and occur even within a single model by stopping the simulation 100 Myr earlier or later. The drop at high gas surface densities in group H is due to the rapid collapse of our highly symmetric initial conditions and subsequent intense SN feedback in the absence of magnetic fields. This is visually demonstrated in Fig.~\ref{fig: visual comparison}, where the ${\Sigma_{\rm{gas,0}}}=80\;M_{\odot}\;\rm{pc}^{-2}$ M and H runs are shown side by side. We discuss this effect in detail in Section~\ref{susbec: exploding disks}.

Comparing to results from other simulations, we find rough agreement with \citet{Rathjen2023} and \citet{Brucy2023}. While \citet{Kim2020} show increased SFR at high values of $\Sigma_{\rm{gas}}$, they vary the external gravitational potential represented by $a_{\rm{DM}}$ and $a_{*}$ such that it increases as $\Sigma_{\rm{gas},0}$ increases, and therefore have an increased gas weight across their simulations. A fair comparison with \citet{Kim2020} is made in terms of gas weight in Section~\ref{sec: prfm}, where we find good agreement with their results. Each of these works uses a different method when computing SFR. \citet{Kim2020} average both SFR and $\Sigma_{\rm{gas}}$ over the period $0.5<t/t_{\rm{orb}}<1.5$, where $t_{\rm{orb}}$ is the orbital time of the disk, which is equal to 220, 110, and 61 Myr, for a mean $\Sigma_{\rm{gas}}$ of 11, 30, and 74 $M_{\odot}\;\rm{pc}^{-2}$. The shorter time used for the computation in the higher $\Sigma_{\rm{gas,0}}$ runs could lead to an artificial deviation from the observed KS relation, if, for example, a large burst of star formation happens to dominate the SFR in the sampled orbital time. \citet{Rathjen2023} average both gas surface density and SFR over the entire simulation run time, excluding the initial phase of turbulent driving, again giving times ranging from 250.5 Myr down to 114.6 Myr for the highest $\Sigma_{\rm{gas,0}}$ models, whose SFR is dominated by the initial burst of SFR. \citet{Brucy2023} average the mass accreted onto sink particles over the period it takes their simulation to go from 97\% to 60\% of its initial gas mass, where gas is lost either via outflows or accretion onto sinks. This means that for the largest $\Sigma_{\rm{gas,0}}$ runs the period over which they calculate their SFR is $\sim 100$ Myr for their TURB models. While these models differ from ours in several aspects, the agreement in the resulting KS relation is encouraging.

\begin{figure}	
	\centering
	\centerline{\includegraphics[trim={0 0.5cm 0 0},clip,width=1\linewidth]{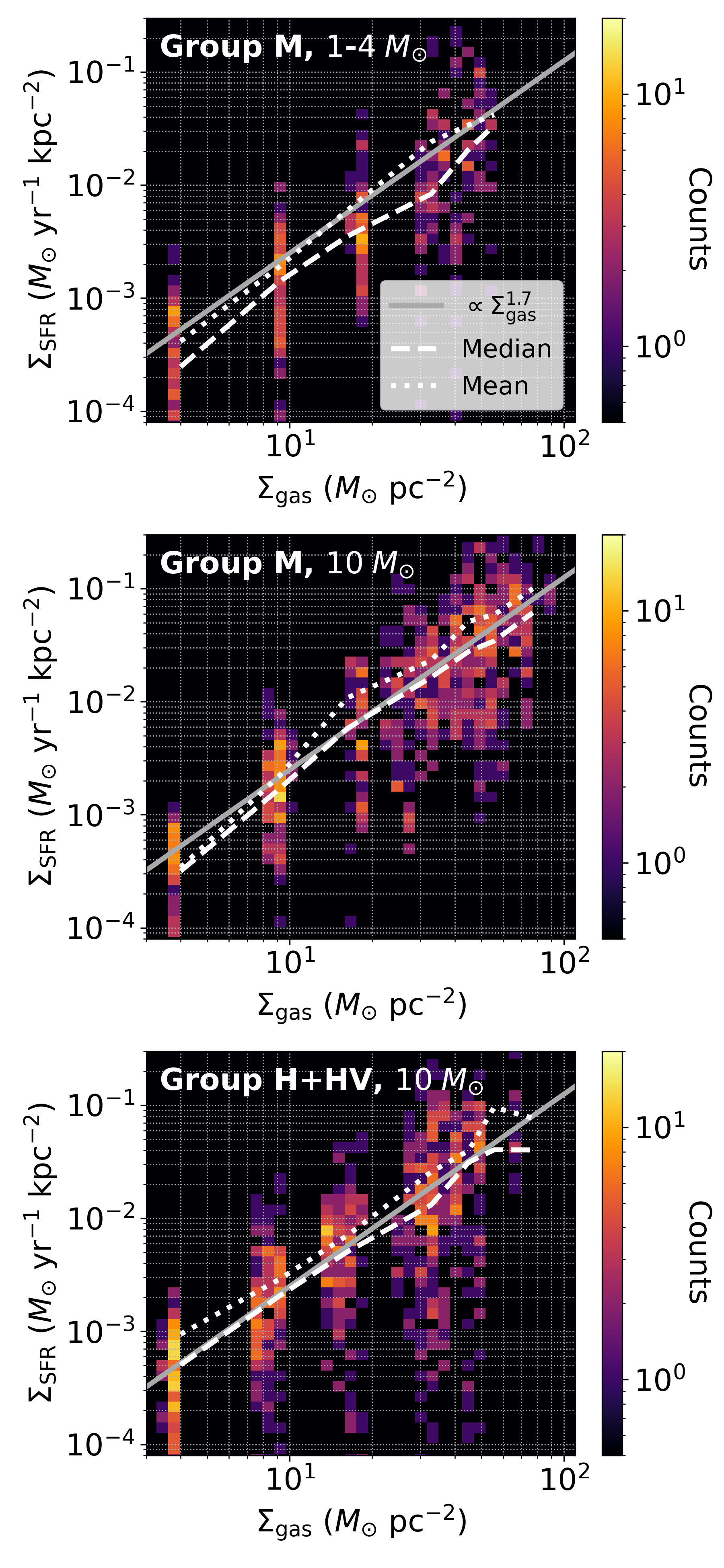}}
	\caption{
Time-resolved KS relation for our different simulation groups. We also plot the median and mean values of $\Sigma_{\rm{SFR}}$ binned along the $\Sigma_{\rm{gas}}$ axis, as well as a power law of the form $\Sigma_{\rm{gas}}^{1.7}$ for visual reference.
		}
		\label{fig: KS resolved}
\end{figure}

In Fig.~\ref{fig: KS resolved}, we bin individual simulation snapshots taken at 1 Myr intervals in $\Sigma_{\rm{gas}}$--$\Sigma_{\rm{SFR}}$ space to create a 2D histogram version of the KS relation. For this analysis, we use results from groups M and H, substituting the runs with $\Sigma_{\mathrm{gas},0}\geq 40\;M_{\odot}\;\mathrm{pc}^{-2}$ with runs from group HV. We also present a power law of the form $\Sigma_{\mathrm{gas}}\propto \Sigma_{\mathrm{SFR}}^{1.7}$ normalized to the solar neighborhood value, for reference. Upon visual inspection, the three panels show a similar KS relation, with the H and HV runs presenting perhaps a larger scatter in $\Sigma_{\mathrm{SFR}}$. As we show in Appendix~A, the large scale vertical pressure support against the disk weight is mostly provided by SN-driven turbulence. In addition, the turbulent pressure we find at a given SFR doesn't depend strongly on the inclusion of magnetic fields. Thus, the self-regulation of star formation by SN feedback results in a similar SFR regardless of whether or not magnetic fields are included. The effect of magnetic fields on the scatter in the KS relation is further discussed in Section~\ref{sec: B fields}.

\subsection{Vertical and Dynamical Equilibrium}
\label{sec: prfm}
In work by \citet{Ostriker2010} and \citet[][hereafter \citetalias{Ostriker2022}]{Ostriker2022}, the formalism for pressure-regulated, feedback-modulated (PRFM) star-formation was established. In this picture, the mid-plane gas weight $\mathcal{W}$ is balanced by the total mid-plane gas pressure $P_{\rm{tot}}$, when averaged over time. In addition, the total pressure, and its magnetic, thermal, and turbulent components, can be related to the star formation rate surface density by the feedback yield $\Upsilon$, defined as

\begin{equation}
    \Upsilon_i\equiv \frac{P_i}{\Sigma_{\rm{SFR}}},
\end{equation}
where the index $i$ represents the relevant pressure component (i.e., magnetic, thermal, turbulent, or the sum of all three). \citetalias{Ostriker2022} showed that $\mathcal{W}$ and $P_{\rm{tot}}$ 
are well balanced in their high gas surface density MHD simulations. In addition, they show that the dynamical equilibrium pressure, defined as 
\begin{equation}
    P_{\rm{DE}}\equiv \frac{\pi G \Sigma_{\rm{gas}}}{2}+\Sigma_{\rm{gas}} \left(2 G \rho_{\rm{sd}} \right)^{1/2} \sigma_{\rm{eff}},
\end{equation}
where $\rho_{\rm{sd}}$ is the mid-plane mass density of the stars and dark matter, and $\sigma_{\mathrm{eff}}$ is the effective velocity dispersion, is a good estimator for the total gas weight. The effective velocity dispersion is defined by
\begin{equation}
    \sigma_{\rm{eff}}\equiv \sqrt{c_s^2+\sigma_B^2+\sigma_z^2},
\end{equation}
where $c_s$ is the sound speed, $\sigma_B$ is given by $\sqrt{\left(B^2-2B_z^2\right)/8\pi\rho}$, 
and $\sigma_z$ is the standard deviation of the $z$-component of the gas velocity, all of which taken as the average across the midplane. In addition, the calculation of $\sigma_{\rm{eff}}$ is restricted to the warm and cold ISM, i.e., gas with $T\leq 2\times 10^4$~K, hereafter referred to as two-phase gas. They compare the relation between $\Sigma_{\rm{SFR}}$ and $P_{\rm{DE}}$ in their simulations with observations, and compute the values of $\Upsilon_i$ in their simulations as a function of $P_{\rm{DE}}$. \citet{Kim2024} investigate the effect of metallicity and dust-to-gas ratio on feedback yields using their improved TIGRESS-NCR framework.

\begin{figure}	
	\centering
	\centerline{\includegraphics[trim={0 0.25cm 0 0},clip,width=1\linewidth]{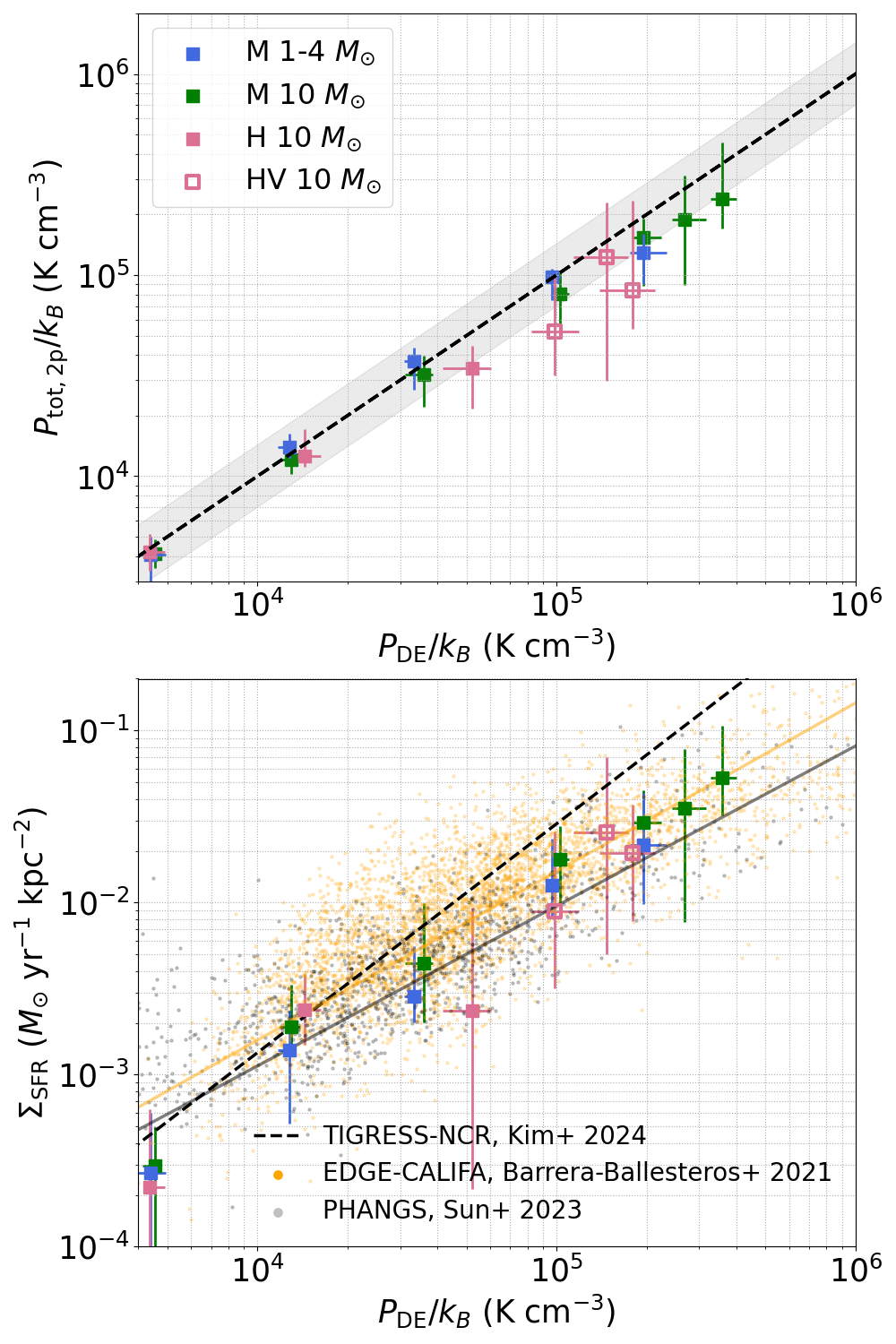}}
	\caption{
Top panel: mid-plane gas pressure in the two-phase gas as a function of the gas weight estimator $P_{\rm{DE}}$ for our different simulations. The dashed line shows the expected 1-to-1 relation, while the gray shaded region shows the $\pm30\%$ range. Bottom panel: star formation rate surface density as a function of $P_{\rm{DE}}$. The dashed line shows the fit to \citet{Kim2024}. Points show observational data from the EDGE-CALIFA \citep{Barrera-Ballesteros2021, Wong2024} and PHANGS surveys \citep{Leroy2021, Sun2023}, and the best fit to their data is shown in solid lines of the same color. Reported simulation results are medians taken over the period of 200-500 Myr, and error bars show the interquartile range. 
		}
		\label{fig: prfm main}
\end{figure}

The top panel of Fig. \ref{fig: prfm main} shows the relation between $P_{\rm{DE}}$ and the total mid-plane pressure in the two-phase gas, as measured by adding up the mid-plane averages of the thermal pressure $P_{\rm{th}}$, vertical magnetic stress $\Pi_B$, and $P_{\rm{turb}}$, when restricted to the two-phase gas. 
We perform volume-weighted averages, using $m/\rho$, where $m$ is the particle mass and $\rho$ is the particle density, as weights.
For the magnetic stress, we consider only the $zz$ (i.e., the vertical) component in the Maxwell stress tensor $\Pi_B=\left(B^2-2B_z^2\right)/8\pi$, while the turbulent pressure is taken to be $\rho\, v_z^2$, where $v_z$ is the $z$-component of the particle velocity. Additionally, we compute the median of the resulting pressure over the period of 200-500 Myr. We find that our simulations fall within 30\% of the one-to-one $P_{\rm{tot,2p}}$--$P_{\rm{DE}}$ relation except for several pure hydrodynamical runs. This is likely due to these models spending a significant time in non-equilibrium states.
The bottom panel of Fig.~\ref{fig: prfm main} shows the relation between $\Sigma_{\rm{SFR}}$ and gas weight as represented by $P_{\rm{DE}}$, as well as data from the PHANGS \citep{Leroy2021, Sun2023} and EDGE-CALIFA \citep{Barrera-Ballesteros2021, Wong2024} surveys. For comparison, we also plot the best fit to simulations presented in \citet{Kim2024}. Our results are in good agreement with observations, with the exception of lower $\Sigma_{\rm{SFR}}$ at low gas weights. 

Our results show a similar $\Sigma_{\rm{SFR}}$--$P_{\rm{DE}}$ slope to the \citet{Kim2024} fit, but an overall lower normalization. This difference is likely due to the resulting SN feedback, as it dominates the overall feedback yield. Several differences between our framework and the TIGRESS model could lead to this offset. The different choice of hydrodynamics code could change the efficiency of supernova bubble expansion. In addition, the different implementation of star particles and the inclusion of differential rotation in TIGRESS could reduce the clustering of stars, leading to subsequently less clustered SNe and affecting the resulting SN feedback.

\subsection{Effect of Magnetic Fields}

\label{sec: B fields}

\subsubsection{Disk Stabilization}
\label{susbec: exploding disks}
Fig.~\ref{fig: KS overview} shows a drop in the KS relation for high gas surface densities in group H. The top row of Fig.~\ref{fig: visual comparison} shows a sample of snapshots from the H80 run. For almost 500 Myr, the gas never forms a quasi-steady disk structure, and is therefore prevented from forming stars. This is a result of our initial conditions, which are symmetrical in the $x$- and $y$- directions. Following the initial collapse of the disk, star formation peaks simultaneously across the box mid-plane, and the subsequent SN feedback is so intense that almost half of the gas is driven out of the box. The remaining gas acquires such a high velocity that it oscillates about the box mid-plane, seemingly beginning to form a disk-like structure at $t=400$~Myr. 

It has been demonstrated that this initial burst of star formation and subsequent feedback can be moderated by a combination of magnetic pressure and artificially driving turbulence for an initial time period until stars begin to form \citep[see, e.g.,][]{Kim2017, Rathjen2023, Brucy2023}. We do not externally drive turbulence in any of our simulations. In contrast to the H80 run, in the M80 run, shown in the bottom row of Fig.~\ref{fig: visual comparison}, magnetic fields slow down the initial collapse of the disk, and effectively create inhomogeneities in the mid-plane. This inhibits the initial extreme burst of star formation, allowing a quasi-steady disk to form. Thus, simulating a kpc-scale ISM without magnetic fields requires mitigation of this initial burst. We find that setting the initial gas distribution as we do for group HV is sufficient to achieve a successful pure hydrodynamical simulation up to $\Sigma_{\rm{gas,0}}=80\;M_{\odot}\;\rm{pc}^{-2}$. This allows us to directly compare between hydrodynamical and MHD simulations in the high column density regime.

\subsubsection{Magnetic Support}
    
\begin{figure}	
	\centering
    \centerline{\includegraphics[trim={0 0.5cm 0 0},clip,width=1\linewidth]{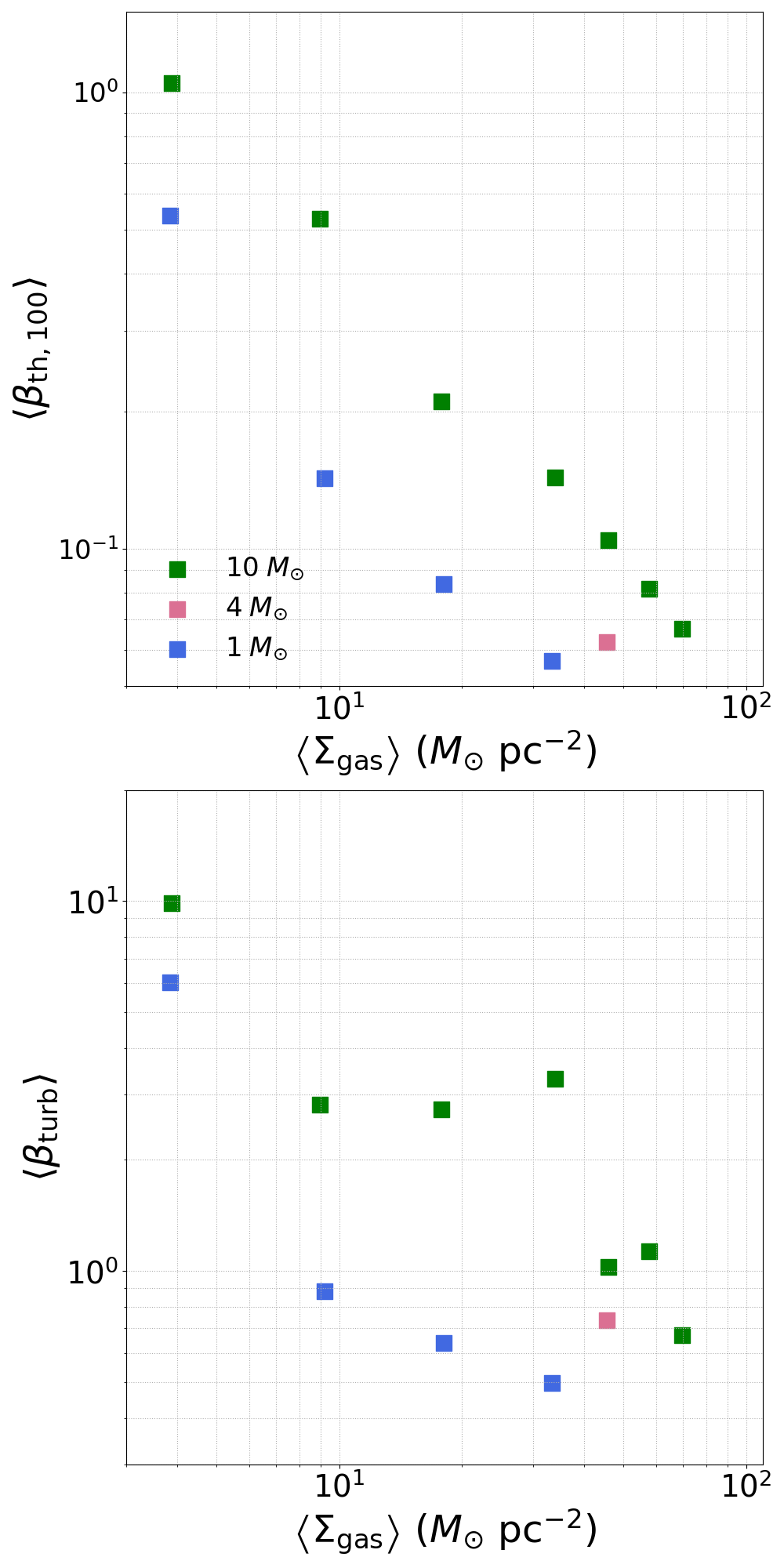}}
	\caption{
Top panel: $\beta_{\mathrm{th},100}$ in our M models, defined as the mass-weighted plasma $\beta$ parameter averaged over gas particles with volume density $n>100\;\rm{cm}^{-3}$, as a function of 
the time averaged gas surface density $\left<\Sigma_{\rm{gas}}\right>$. 
Bottom panel: same as top panel, but for $\beta_{\mathrm{turb}}$, the ratio between turbulent and magnetic pressure measured on 10 pc scales.
		}
		\label{fig: beta vs SG}
\end{figure}

In addition to reducing the strength of SN feedback that dominates in the first $\sim 200$ Myr of our simulations, magnetic fields also affect the evolution and structure of the ISM. To compare the importance of the magnetic and thermal pressure in star-forming gas, we define $\beta_{\rm{th},100}$ to be the mass-weighted average of the plasma $\beta$ parameter (the ratio of the thermal to magnetic pressures), averaged over gas particles with volume density $n>100\;\mathrm{cm}^{-3}$. The top panel of Fig.~\ref{fig: beta vs SG} shows the relationship between the time average of $\beta_{\rm{th},100}$ and the time-averaged $\Sigma_{\rm gas}$ taken over a simulation time of 200--500 Myr, demonstrating a clear decreasing trend. In the dense gas, which tends to be cold, magnetic pressure is non-negligible already at low 
$\left<\Sigma_{\rm{gas}}\right>$
and becomes increasingly dominant as  
$\left<\Sigma_{\rm{gas}}\right>$
increases. In addition, there is a systematic trend with resolution, where $\beta_{\mathrm{th},100}$ is lower for higher resolution. 

Repeating this analysis for the turbulent pressure component, we define $\beta_{\rm{turb}}$ as the ratio of the turbulent to magnetic pressures in the box mid-plane. To calculate $\beta_{\rm{turb}}$, we first partition the box mid-plane, defined by $\left|z\right|<10\;\rm{pc}$, into cubes with a side length of 10 pc. For each cube, we calculate the mass-weighted vertical components of the magnetic and turbulent pressures as defined in Section~\ref{sec: prfm}. We then calculate $\beta_{\rm{turb}}$ by taking the mass-weighted average of the ratio $\left(P_{\rm turb}/\Pi_B\right)$ across all of these sub-volumes over a simulation time of 200-500 Myr. We do not restrict $\beta_{\rm{turb}}$ to dense cubes, as very few regions would fulfill the $n>100$ criterion and the results would become quite noisy due to averaging over a small number of regions. Similar to $\beta_{\rm{th},100}$, the bottom panel of Fig.~\ref{fig: beta vs SG} shows that $\beta_{\rm{turb}}$  decreases with increasing $\left<\Sigma_{\rm{gas}}\right>$ and with improved resolution. This method of comparing turbulent and magnetic support is more precise when discussing turbulent pressure support on the scale of individual clouds. The (volume-weighted) average of $\sigma_{\rm{z}}$ over the entire midplane is still the appropriate quantity when discussing vertical pressure support and dynamical equilibrium on the scale of the entire disk (see Appendix A).
Both of these results are consistent with magnetic pressure playing an important role in supporting the cold ISM on cloud scales against gravitational collapse, with growing importance as the gas surface density increases. 

Since magnetic pressure in star-forming gas in our simulations becomes increasingly important as the gas surface density increases, we can expect star-forming clouds to collapse faster in high $\Sigma_{\rm{gas},0}$ runs when magnetic fields are turned off. 
However, when averaged over large timescales, the SFR is not affected. This is demonstrated in Section~\ref{sec: star formation} for simulations that  successfully form quasi-steady disks, which indeed occurs in all of our models excluding H80. In these cases, feedback-induced self-regulation always drives our simulations towards the observed KS relation. 

\begin{figure}	
	\centering
	\centerline{\includegraphics[trim={0 0.5cm 0 0},clip,width=1\linewidth]{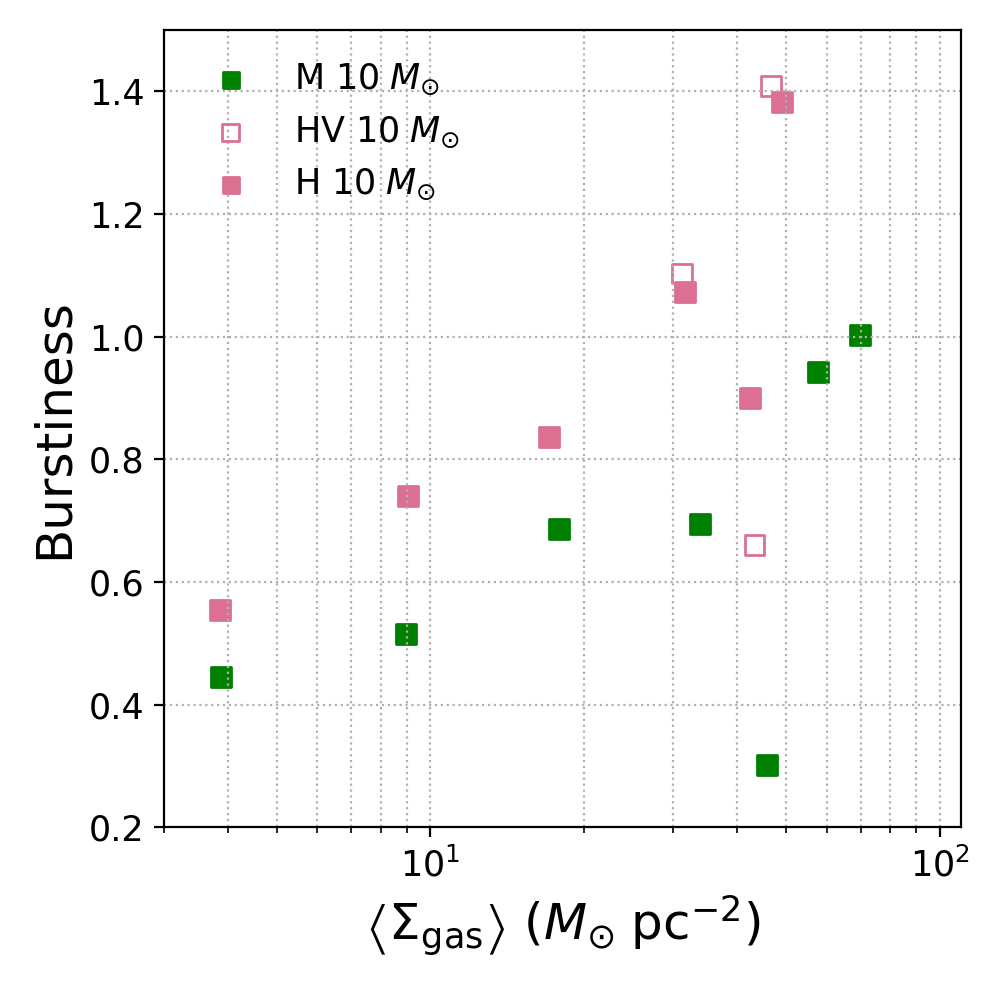}}
	\caption{
Burstiness parameter, defined as $\frac{\sigma(\mathrm{SFR})}{<\mathrm{SFR}>}$, as a function of the time average $\left<\Sigma_{\rm{gas}}\right>$, for our low-resolution models.
		}
		\label{fig: burstiness}
\end{figure}

\begin{figure}	
	\centering
	\centerline{\includegraphics[trim={0 0.2cm 0 0},clip,width=1\linewidth]{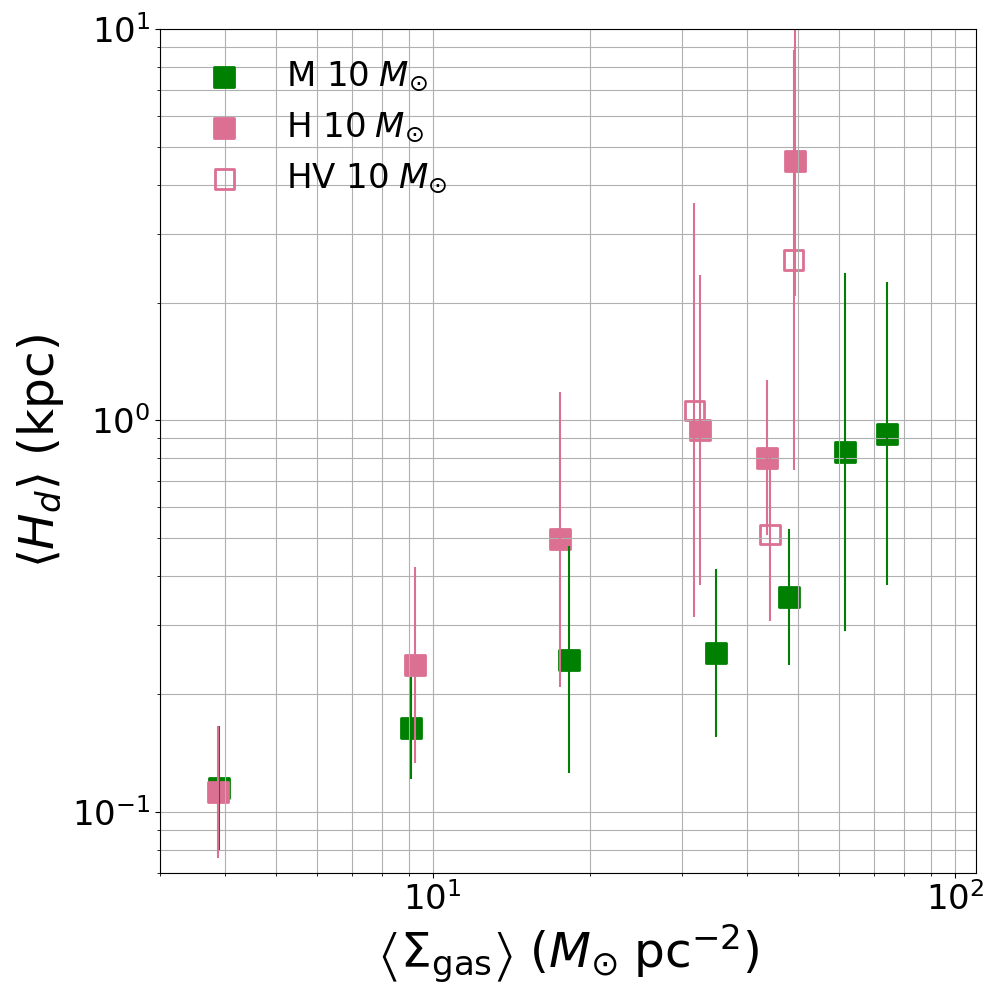}}
	\caption{
Effective disk-height as a function of the time average $\left<\Sigma_{\rm{gas}}\right>$ for our $10\;M_{\odot}$ models in Groups M, H, and HV. Values reported are averages taken over the period 200-500 Myr, error bars present the standard deviation in logarithmic space.
		}
		\label{fig: scale height}
\end{figure}

Fig.~\ref{fig: burstiness} shows the burstiness parameter for our low-resolution models, defined following \citet{Zhang2024} as
\begin{equation}
    \mathrm{Burstiness}=\frac{\sigma(\mathrm{SFR})}{\left<\mathrm{SFR}\right>}
\end{equation}
where SFR is computed as is described in Section~\ref{sec: star formation}, and $\left<\rm{SFR}\right>$ is averaged over the simulation time between 200-500 Myr, and $\sigma\left( \rm{SFR} \right)$ is the standard deviation of the SFR in that period.
We compare groups M, H, and HV, and find that simulations from groups H and V are burstier than group M, and that, except for $\Sigma_{\rm{gas},0}=60 \; M_{\odot} \; \rm{pc}^{-2}$  ($\left< \Sigma_{\rm{gas}} \right>\approx40-50\;M_{\odot}\;\rm{pc}^{-2}$), generally follow a trend where higher surface density leads to burstier star formation. Additionally, the difference in burstiness between groups H/HV and group M increases with gas surface density.
Fig.~\ref{fig: scale height} shows the time-averaged effective disk height, $H_{d}$, defined as the distance from the mid-plane in which 50\% of the gas mass is contained. We can see that there is a general increasing trend in disk height with gas surface density, and that the disk height is larger at a given gas surface density in group H and HV, compared to group M. Moreover, the difference is larger for higher $\left<\Sigma_{\rm{gas}}\right>$.
The increase in burstiness and disk height can be explained by the fact that in the absence of magnetic fields, the faster collapse can lead to more stars forming from a single gas cloud before being dispersed by eventual SN feedback. The resulting SN feedback from a larger cluster of stars will then be stronger and more effective, and will drive more gas further out of the disk mid-plane. This, in turn, increases the effective disk height, as well as the gas outflow rate. We defer a full investigation of the effect of gas surface density and magnetic fields on the energy- and mass-loading factors to a future study.

\subsubsection{ISM Phase Distribution}

\begin{figure}	
	\centering
    \centerline{\includegraphics[trim={0 0 0 0},clip,width=0.9\linewidth]{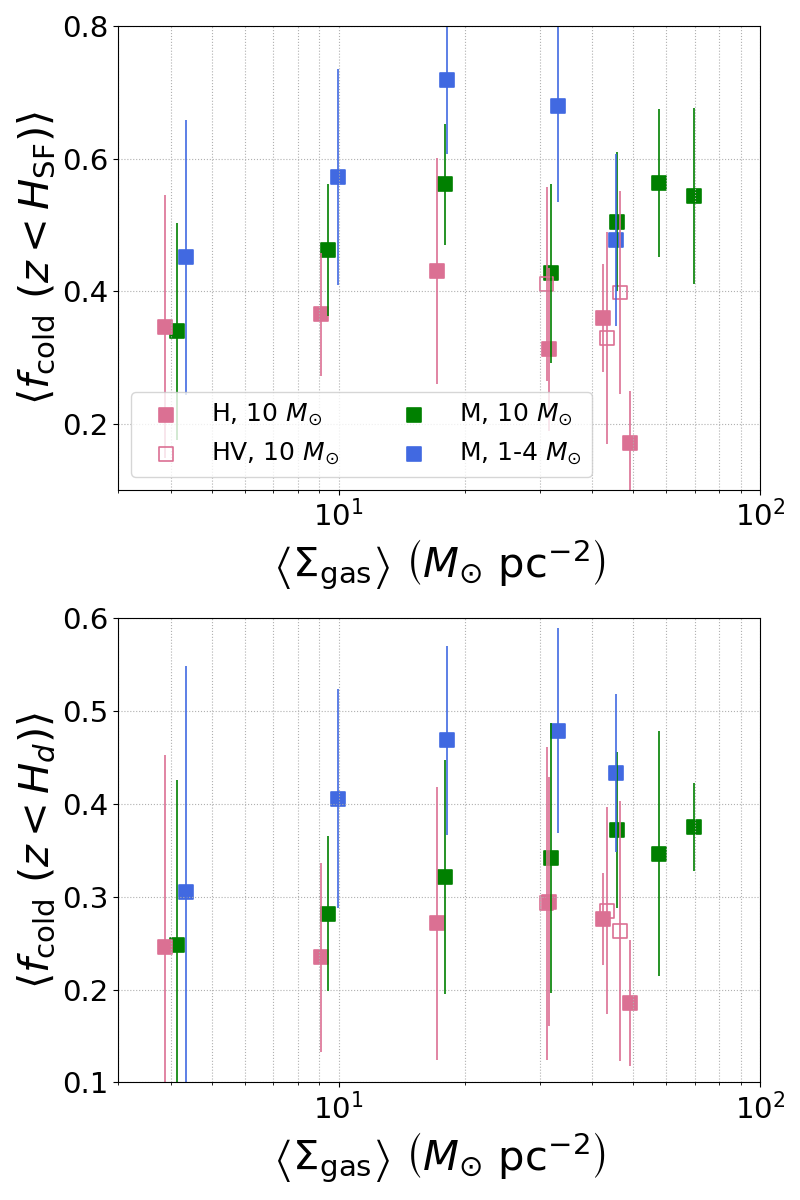}}
	\caption{
Top panel: mass fraction in the cold ($T<300\ \rm{K}$) gas in the region $\left| z \right| <H_{\rm{SF}}$, defined as the star formation weighted average height, as a function of mean gas surface density. Bottom panel: same as top panel, but for $\left| z \right| <H_{d}$, the half-mass height of the simulation. $H_{\rm{SF}}$, $H_d$, and the cold gas mass fractions are computed by averaging each simulation over the time period of 200-500 Myr. Error bars show the standard deviation. Points with $\left<\Sigma_{\rm{gas}}\right>\leq 10\ M_{\odot} \ \rm{pc}^{-2}$ are shifted horizontally by small values for visual clarity.
		}
		\label{fig: phases}
\end{figure}

Fig.~\ref{fig: phases} shows the ISM cold gas mass fraction in our models as a function of $\left<\Sigma_{\rm{gas}}\right>$. We compute the cold gas fraction both for gas particles whose $z$-coordinate is within $H_d$ (the half-mass height of the disk) and within $H_{\rm{SF}}$ (the SFR-weighted average midplane distance). We designate a gas particle as cold if its temperature is less than 300 K. The remaining mass is mainly warm neutral gas with $T\sim 10^4\;\rm{K}$, with a contribution of order a few \% from hot ($T\sim 10^6\;\rm{K}$) and photoionized gas. 

For both definitions of $f_{\rm{cold}}$, we find that models from group M show systematically higher values compared to groups H and HV. Averaging the relative difference in $f_{\rm cold}$ between groups H and M with equal $\Sigma_{\rm gas,0}$ and at low resolution (and excluding models H80 and M80), we find an increase of 26\% and 18\% for gas within $H_{\rm SF}$ and $H_{\rm d}$, respectively. For gas within $\left| z \right|<H_d$, we also find an increase in $f_{\rm{cold}}$ with increasing $\left< \Sigma_{\rm{gas}}\right>$. In addition, we find that higher resolution models show higher cold gas fractions. This is possibly due to the high-resolution models resolving more of the ISM dense substructure, allowing it to cool more efficiently, but a detailed investigation of this resolution effect is deferred to future work. We also see that group simulation H80 ($\left< \Sigma_{\rm{gas}}\right>\approx 40\;M_{\odot}\ \rm{pc}^{-2}$) shows lower values of $f_{\rm{cold}}$. This can be explained simply by this simulation never establishing a quasi-stable disk and therefore lacking dense and cold gas.

The reason for the increase in the cold gas fraction in group M compared with groups H and HV is once again the magnetic pressure, which becomes increasingly important as the gas surface density increases. As it slows down the collapse of cold, star-forming, gas clouds, it allows cold gas to survive for a longer period before the onset of feedback that consequently destroys the cloud. This means that correctly modeling magnetic fields in the ISM is crucial for the study of the cold ISM, and particularly the study of emission lines that trace cold gas, such as CO rotational lines, [$\text{C\,\small{I}}$] 640 and 390 $\mu$m, and [$\text{C\,\small{II}}$] 158 $\mu$m.
Thus, modeling these lines to study the relationship between emission, cold or molecular gas mass, and star formation, is likely strongly affected by the inclusion or neglect of magnetic fields.

\section{Steady State Magnetic Field}

\label{section: B fields tests}

In this section, we demonstrate that a steady-state magnetic field emerges in our simulation, whose value is determined by the gas surface density. 

\subsection{\(\mathbf{B}\)-field Independence of Initial Conditions}

\begin{figure}	
	\centering
	\centerline{\includegraphics[trim={0 0 0 0},clip,width=1\linewidth]{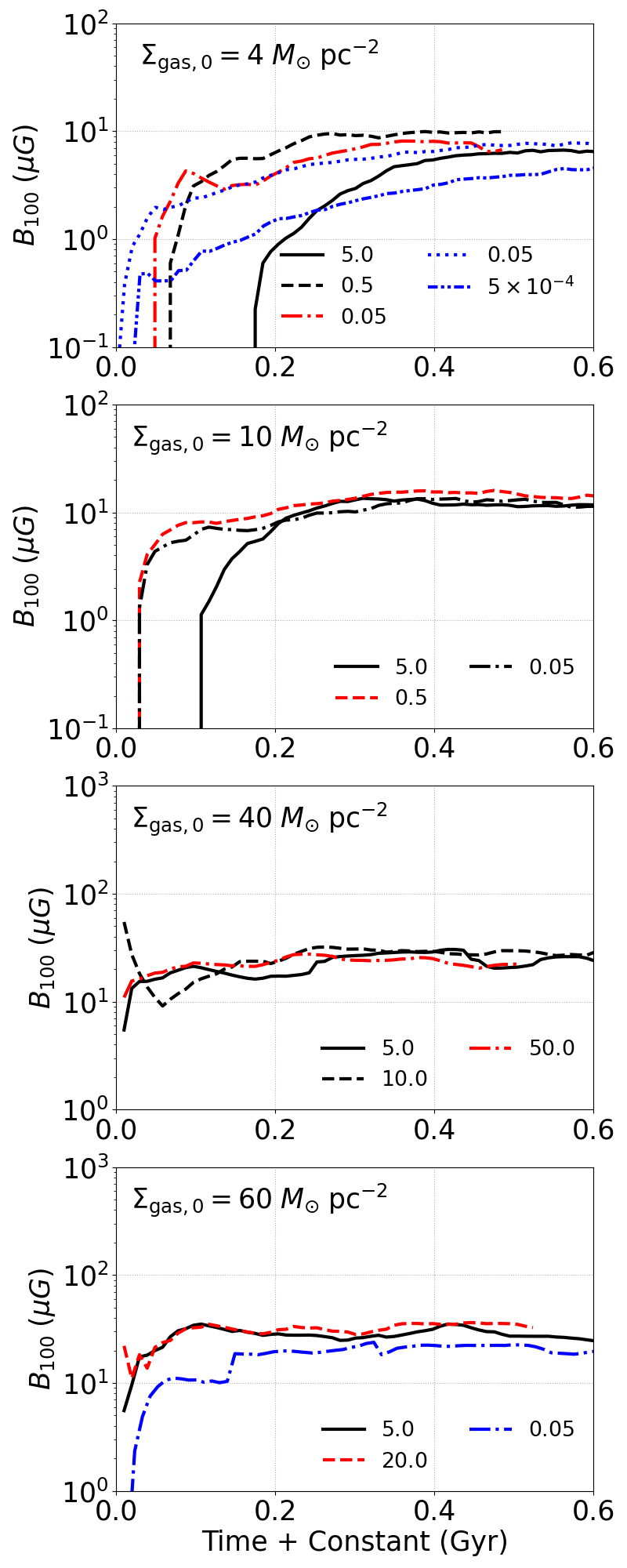}}
	\caption{
Mass-weighted mean $B_{100}$ as a function of time. Each panel shows simulations with a single value of $\Sigma_{\mathrm{gas,0}}$ indicated in the upper left corner of each panel. The value of $B_0$ is given in the legend in $\mu$G. $B_{100}$ is smoothed using a 200 Myr window in order to visually demonstrate its convergence. Red lines denote our group M runs, blue lines denote our group MV runs. 
		}
		\label{fig: B vs time test}
\end{figure}

Fig.~\ref{fig: B vs time test} shows the mass-weighted magnetic field in the dense gas ($n>100\;\mathrm{cm}^{-3}$) gas, denoted as $B_{100}$, as a function of time for an additional set of simulations listed in Table~\ref{table: B tests}. These simulations were run with magnetic fields turned on and with an identical setup to Group M (black and red curves) or MV (blue curves), the only difference being the value of $B_0$. These simulations were run using a resolution of $m_{\rm{g}}=10\;M_{\odot}$. We also apply a moving average with a window of 200 Myr to $B_{100}$ as its fluctuations in time (of order 1 dex) qualitatively follow the fluctuations in SFR, and obscure its convergence. The red lines mark our fiducial models which we use for the analysis in the rest of the paper (group M). In all cases, we find that variations in the initial magnetic field by up to  4 orders of magnitude, and the initial orientation of the magnetic field, all result in variations in the steady-state value of $B_{100}$ by less than a factor of $\sim 2$. In the $\Sigma_{\rm{gas},0}=4\;M_{\odot}\;\rm{pc}^{-2}$ case, we find that this convergence takes a longer time, of the order of a few 100 Myr. If the setting of $B_{100}$ is related to SN driven turbulence, then we can associate this slower convergence of $B_{100}$ to the lower SFR in $\Sigma_{\rm{gas},0}=4\;M_{\odot}\;\rm{pc}^{-2}$ models. Our $\Sigma_{\rm{gas},0}=10\;M_{\odot}\;\rm{pc}^{-2}$ models take $\sim 100$ Myr to stabilize, while $\Sigma_{\rm{gas},0}=40\rm{\;and\;60}\;M_{\odot}\;\rm{pc}^{-2}$ achieve convergence almost instantly, all in line with the convergence time being related to SFR. 
In Table~\ref{table: B tests} we show $B_0$, as well as the values of $B_{100}$ and $\Sigma_{\rm{SFR}}$, both averaged over the final 200 Myr of the simulation, for the different tests shown in Fig.~\ref{fig: B vs time test}. We see that even when $B_0$ is varied by orders of magnitude, $B_{100}$ reaches the same steady state value of order $\sim 10$~$\mu$G, to within a factor of $\sim 2$, and the SFR is also weakly affected.
\begin{table}
\centering
\begin{threeparttable}

\centering
\begin{tabular}{l l l l}
\hline \hline 
\\ [-1.5ex] $\Sigma_{\rm{gas,0}}\;(\rm{a})$ & $B_0$ (b) & $B_{100,\rm{final}}$ (c)& $\Sigma_{\rm{SFR}} \;(\rm{d})$ \\ [0.5ex] 
\hline

\\[-3ex]4 & 5 & 7.62 & 2.53
\\[0.5ex] & 0.5 & 9.88 & 2.58
\\[0.5ex] & 0.05 & 6.73 & 3.70
\\[0.5ex] *& 0.05 & 6.48 & 5.04 
\\[0.5ex] *& $5\times 10^{-4}$ & 4.80 & 5.85 \\ [0.5ex]

\hline
\\[-3ex]10& 5 & 15.3 & 18.9
\\[0.5ex]& 0.5 & 13.6 & 23.1
\\[0.5ex]& 0.05 & 11.2 & 29.1 \\ [0.5ex]
\hline
\\[-3ex]40& 50 & 22.3 & 178
\\[0.5ex]& 10 & 21.8 & 238
\\[0.5ex]& 5 & 14.9 & 168 \\ [0.5ex]
\hline
\\[-3ex]60& 20 & 32.6 & 334
\\[0.5ex]& 5 & 25.8 & 176
\\[0.5ex]* & 0.05 & 19.9 & 194

\\[0.5ex]
\hline

\end{tabular}
\caption{List of our simulation runs with different values of $B_0$. Simulations use our group M setup, except for rows marked by an asterisk which use our group MV setup. These simulations were all run using a resolution of $m_{\rm{g}}=10\;M_{\odot}$. Values are given in units of (a) $M_{\odot}\;\rm{pc}^{-2}$, (b) $\mu \rm{G}$, (c) $\mu \rm{G}$, (d) $10^{-4} M_{\odot}\;\rm{kpc}^{-2}\;\rm{yr}^{-1}$.}
\label{table: B tests}
\end{threeparttable}
\end{table}

\subsection{\(\mathbf{B}\) -- \(\Sigma_\mathrm{gas}\) Relation}

\begin{figure}	
	\centering
	\centerline{\includegraphics[trim={0 0 0 0},clip,width=1\linewidth]{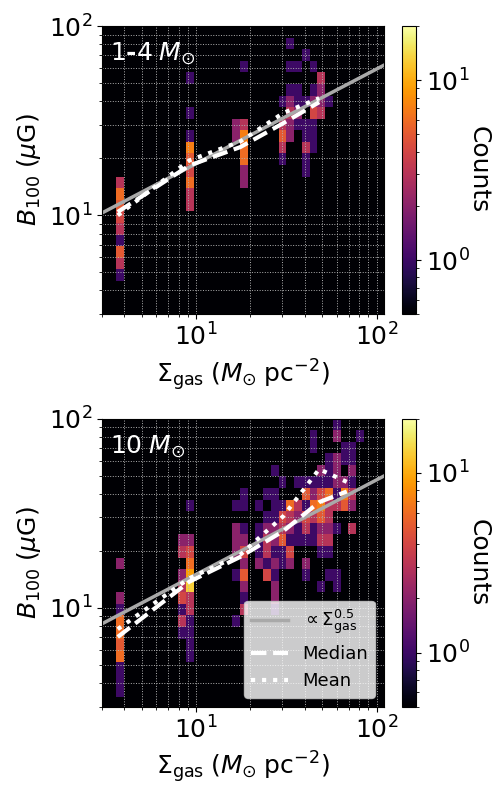}}
	\caption{
Time-resolved $B_{100}$--$\Sigma_{\rm{gas}}$ relation for simulations in group M at a resolution of $1$-$4\;M_{\odot}$ (top panel) and $10\;M_{\odot}$ (bottom panel).
		}
		\label{fig: B-SG resolved}
\end{figure}

Given that our tests show that a steady-state $B_{100}$ emerges from our simulations independently of our initial conditions, we can investigate the relationship between the $B_{100}$ and $\Sigma_{\rm{gas}}$. Fig.~\ref{fig: B-SG resolved} shows the $B_{100}-\Sigma_{\rm{gas}}$ relation in the form of a 2-dimensional histogram, as well as the binned mean and median values of $B_{100}$, and a power law of the form $B\propto {\Sigma}_{\rm{gas}}^{0.5}$.

\begin{figure}	
	\centering
	\centerline{\includegraphics[trim={0 0.5cm 0 0},clip,width=1\linewidth]{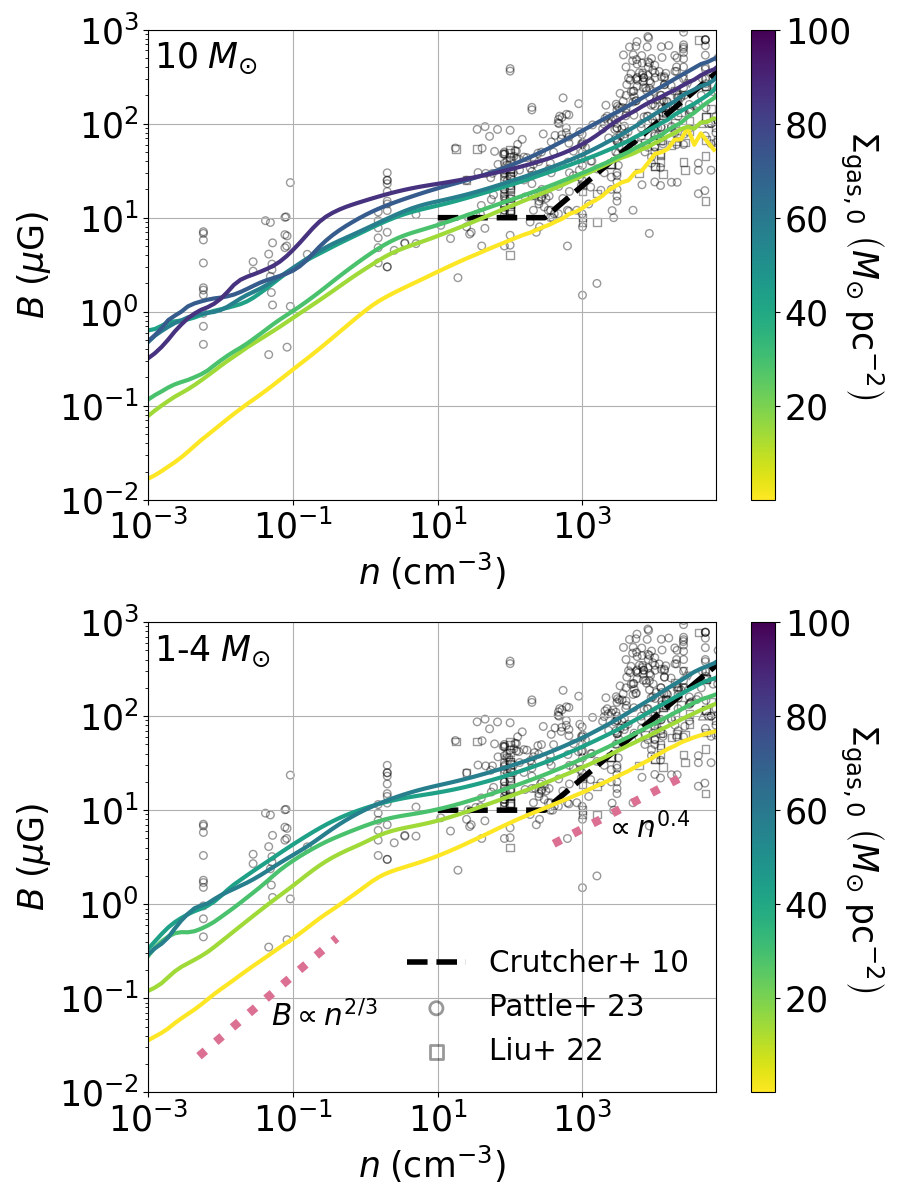}}
	\caption{
Mean magnetic field as a function of gas density for simulations in group M. Mass resolution is specified in the upper left of each panel, and $\Sigma_{\rm{gas},0}$ is specified by the color of the curve. In addition, we plot the broken power law fit to observations of \citet{Crutcher2010}, and the data points from \citet{Liu2022} and \citet{Pattle2023}. We also plot power laws with a slope of 0.6 and 0.4 for reference.
		}
		\label{fig: B-n relation}
\end{figure}

Fig.~\ref{fig: B-n relation} shows the average relation between the magnetic field strength and gas density for group M. We calculate the relation by taking simulation data between times 200 and 500 Myr, binning particles according to density, and taking the mass-weighted magnetic field strength in each bin. We find that at high densities, the average magnetic field is higher at a given density for higher values of $\Sigma_{\rm{gas},0}$. The $B$--$n$ relation also follows a power law with a slope of $\sim0.4$, close to the slope predicted from theory and simulations of the collapse of strongly magnetized dense clouds \citep{Padoan1999}. In the low-density gas, we find that the $B$--$n$ relation roughly follows a power law dependence with a slope of 2/3, as is expected in the flux-freezing limit. We do see some of the curves intersecting at low densities, perhaps due to a dependence of the magnetic field in the diffuse phase on our initial conditions. This should not affect the dynamics of the simulation, as magnetic pressure is negligible compared to the thermal pressure at these low densities.

The fact that at a given density, the average $B$-field is larger for larger values of $\Sigma_{\rm{gas,0}}$ demonstrates that the physical mechanism that sets the $B_{100}$--$\Sigma_{\rm{gas}}$ is related to the increase in $\Sigma_{\rm gas ,0}$ itself.
A natural explanation is that as the gas surface density increases, so does the star-formation rate surface density and, in turn, the SN rate surface density. This leads to increased turbulence, which then amplifies the magnetic field to higher values through the small-scale dynamo effect.

\section{Summary}
\label{sec: summary}

We present a new suite of high-resolution (1-10 $M_{\odot}$, $\sim$0.2-2 pc) 3D simulations of a galactic patch, with a resolved cold ISM, star-by-star treatment of star formation, SN and photoionization feedback. Our models span a wide range in gas surface density, and include both MHD and pure hydrodynamical simulations at high gas surface densities, running for 500 Myr or more. This allows us to capture the evolution and importance of magnetic fields in this column density regime.

\begin{enumerate}[leftmargin=*, noitemsep]
    
    \item Our KS relation agrees with observations and other theoretical work, with a power law index in the 1.4-1.7 range.
    \item Our simulations are in vertical dynamical equilibrium. The relation between $\Sigma_{\rm{SFR}}$ and total gas weight in our simulations is in good agreement with observations and agrees the TIGRESS-NCR simulations \citep{Kim2024}.
    \item We find that magnetic fields play an important role at the high gas surface density regime ($\Sigma_{\rm{gas,0}}\gtrsim40\;M_{\odot}\;\rm{pc}^{-2}$), by reducing the strong fluctuations in SFR which are prevalent in the pure-hydrodynamical models, but without changing the SFR when averaged over long times.
    \item As a result, simulations with magnetic fields show up to a 40\% increase in cold gas fraction and a reduction in disk scale height by up to a factor of 2.
    \item Magnetic fields help mitigate the initial strong burst of star formation known to occur in galactic patch simulations. We demonstrate that this mitigation can also be achieved by using more turbulent initial conditions.
    \item We find that the magnetic field strength in the dense gas is insensitive to the initial magnetic field strength and configuration, and that a $\Sigma_{\rm{gas}}$-dependent, quasi-steady state magnetic field strength emerges from our simulations, approximately following a power law of the form $B\propto \Sigma_{\rm{gas}}^{0.5}$.
    \item The $B$-$n$ relation in our models shows a power law slope of $2/3$ and 0.4 in the diffuse and dense regimes, respectively.
    The normalization of the relation changes with the gas surface density, in agreement with observations despite the large scatter.
    This points to the magnetic field strength being set by SN-driven small-scale dynamo.
    \end{enumerate}
     Future prospects for analysis based on the GHOSDT simulations are abundant. These include the effect of $\Sigma_{\rm{gas}}$ on chemical abundances, ISM phase structure, outflow statistics, and line emission from atomic and molecular gas. In addition, we now have the capability to compare pure hydrodynamical simulations with their MHD counterparts. Finally, the extension into even higher values of $\Sigma_{\rm{gas}}$ could help shed light on the gas properties of gas-rich starburst galaxies and the high star formation efficiencies observed in the brightest galaxies in the early universe.

We thank Chang-Goo Kim and Eve C. Ostriker for fruitful discussions. We thank Kate Pattle and Jorge K. Barrera-Ballesteros for sharing observational data sets. We thank the referee for their very helpful and constructive comments. This work was supported by the German Science Foundation via DFG/DIP grant STE/1869-2 GE 625/17-1, 
by the National Science and Technology Council (NSTC) of Taiwan under Grant No. NSTC 113-2112-M-002-041-MY3 and National Taiwan University under Grant No. NTU-NFG-114L7444,
by the Center for Computational Astrophysics (CCA) of the Flatiron Institute, and the Mathematics and Physical Sciences (MPS) division of the Simons Foundation, USA.

\bibliography{library}
\bibliographystyle{aasjournal}

\section*{Appendix A - Feedback Yields and Energy Partition}
\label{appendix: feedback yields}

\begin{figure}
	\centering
	\centerline{\includegraphics[trim={0 0.25cm 0 0},clip,width=1\linewidth]{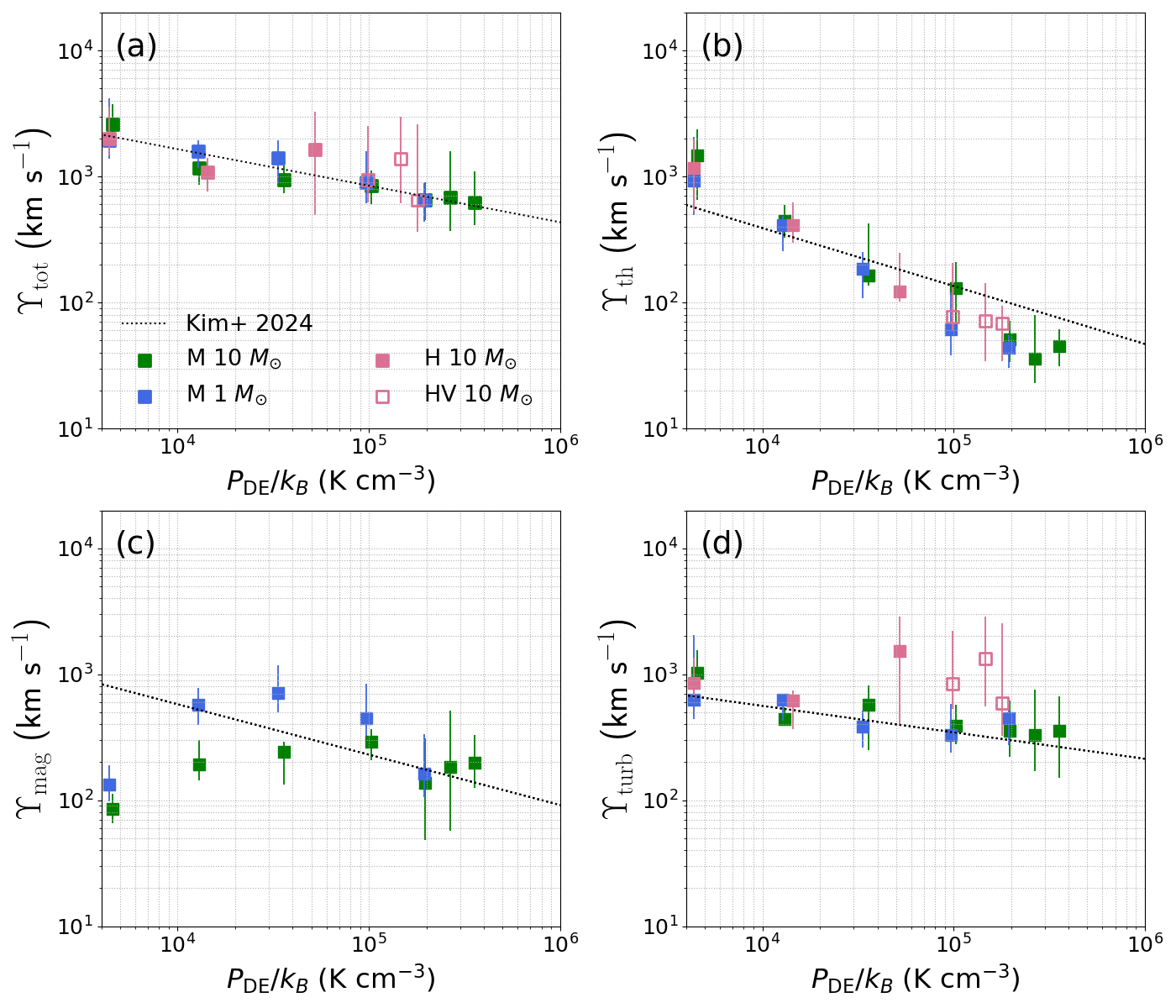}}
	\caption{
Feedback yields as a function of gas weight as estimated by $P_{\rm{DE}}$ for our different simulation groups. Dotted black lines show fits to the results of \citet{Kim2024}. Reported values are medians taken over the period 200-500 Myr and error bars show the interquartile range.
		}
		\label{fig: prfm yields}
\end{figure}

In this appendix we present the feedback yields and compare the kinetic and magnetic energy contributions for our different models. In Fig. \ref{fig: prfm yields} we show the total, thermal, magnetic, and turbulent yields as a function of $P_{\rm{DE}}$, computed by taking the median of the respective yield over the period 200-500 Myr in each simulation. Our total yields agree with the results of \citet{Kim2024}, excluding two pure-hydrodynamical runs. Examining the individual pressure components, we find that at low values of $P_{\rm{DE}}$, the thermal yield is slightly higher in our simulations. At higher values of $P_{\rm{DE}}$, the thermal yield dips below the \citet{Kim2024} fit. The turbulent yields of agree well, except for the four highest $\Sigma_{\rm{gas,0}}$ pure hydrodynamical runs, which are shown in Fig.~\ref{fig: prfm main} to be slightly out of equilibrium. The magnetic yield is comparable to the findings of \citet{Kim2024} (except for the lowest gas surface density runs), although further investigation is required to understand the resolution dependence in our simulations.

\begin{figure}
	\centering
	\centerline{\includegraphics[trim={0 0.25cm 0 0},clip,width=1\linewidth]{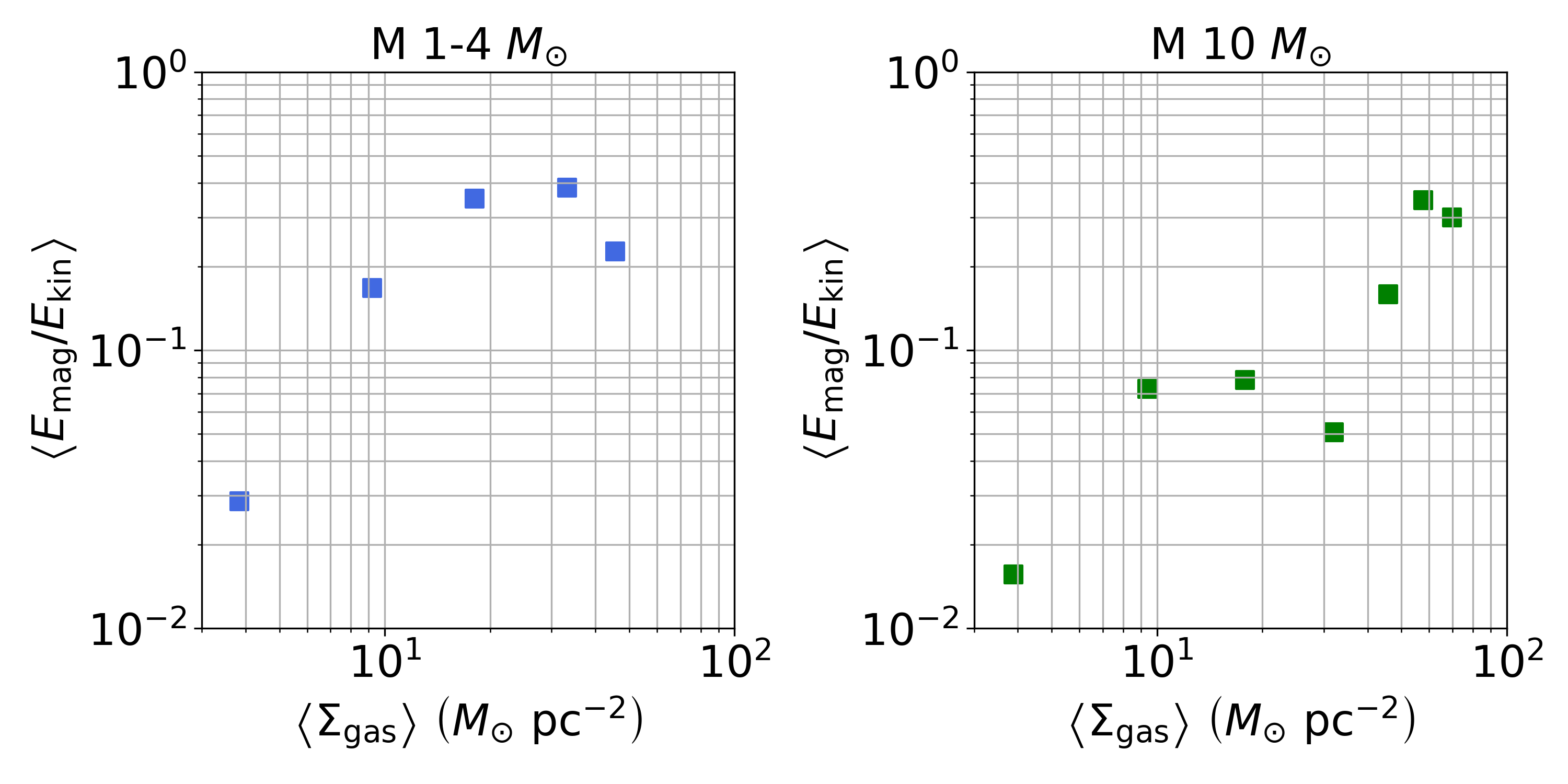}}
	\caption{
The ratio of magnetic and kinetic energy in our entire box averaged over time, as a function of average gas surface density, taken over the period 200-500 Myr. Mass resolution is indicated in the title of each panel.
		}
		\label{fig: turb/mag}
\end{figure}

Fig.~\ref{fig: turb/mag} shows the ratio median ratio between the total magnetic and kinetic energy within 1 kpc of the midplane of our simulation for models in group M. We find ratios as high as 40\%. It could be seen as being in tension with simulations of SN-driven small-scale dynamo on either a galactic scale \citep{Pakmor2017, Rieder2017} or small scale volumes \citep{Gent2021, Gent2023}, where the saturation of the magnetic energy is of a few \% (and up to 20\% in the central 1 kpc in the models of \citet{Pakmor2017}).  While our resulting $B-n$ relation broadly agrees with observed values, the higher-than expected global contribution of the magnetic energy necessitates a more in-depth future study of the small scale dynamo in our simulation. In addition, one must point out that approach of setting a globally constant magnetic field is ill advised, and should be revised in our future simulations, opting for more reasonable setups, e.g. setting the initial magnetic field to represent a constant fraction of the initial thermal energy. This, in turn, would prevent particles at the largest distance from the mid-plane at $t=0$ from carrying unrealistic amounts of magnetic energy.

\section*{Appendix B - Evolution of The Magnetic Field Configuration}

\label{appendix: Bx}

\begin{figure}[h]
	\centering
	\centerline{\includegraphics[trim={0 0 0 0},clip,width=1\linewidth]{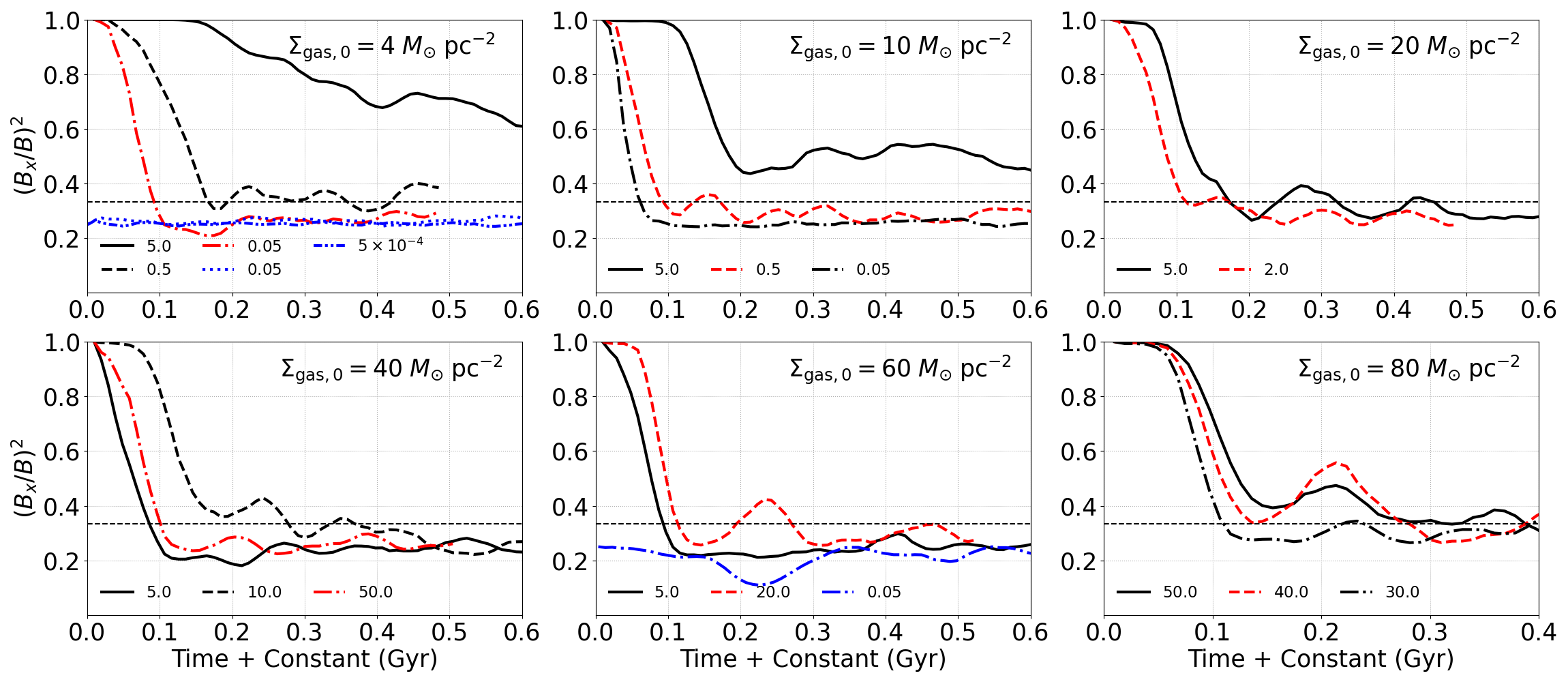}}
	\caption{
Time dependence of the relative strength of the $x$-component of the magnetic field, as represented by $\left(B_x/B\right)^2$. Each panel consists of simulations with $\Sigma_{\rm{gas,0}}$ specified in its top right corner. Red lines are simulations from group M, blue line are simulations from group MV, and black lines have the setup of group M but a different value of $B_0$, specified in the legend. We apply a 50 Myr moving average to the results for easier visualization.
		}
		\label{fig: Bx vs time}
\end{figure}

Our analysis of the magnetic field strength requires that the magnetic field configuration and magnitude arises self-consistently from our simulations without a dependence on initial conditions. In Section~\ref{section: B fields tests} we showed that the magnetic field strength reaches a quasi-steady state. Here we show our magnetic configuration is also independent of initial conditions. Fig. \ref{fig: Bx vs time} shows the mass-weighted average of the value of $\left(B_x/B \right)^2$, where $B_x$ is the $x$-component of the magnetic field, and $B$ is its magnitude. It includes simulations with a mass resolution of $10\;M_{\odot}$ only, with the initial magnetic field listed in the legend of each panel. While the initial value is 1 in group M by construction, it is closer to 1/3 for group MV, and we would expect it to decrease to or remain approximately at that value for us to determine that the initial conditions of the simulation have been erased. Indeed, Fig. \ref{fig: Bx vs time} shows that all of the simulations in group M and HM, respectively in red and blue, reach this desired state. In fact, this holds for a number of other tests simulations where we vary the initial magnetic field. We do find, however, that if the initial magnetic field is high enough (for a given gas surface density), $B_x$ can continue dominating for several hundred Myr, as is demonstrated in the $B_0=5\;\mu G$ runs for the two lowest gas surface densities. This is therefore a test that MHD galactic patch simulations should pass in order to avoid an initial conditions-dominated magnetic field.

\section*{Appendix C - Maintaining $\nabla \cdot \bs{B}\approx 0$}
\begin{figure}[h]
	\centering
	\centerline{\includegraphics[trim={0 0 0 0},clip,width=1\linewidth]{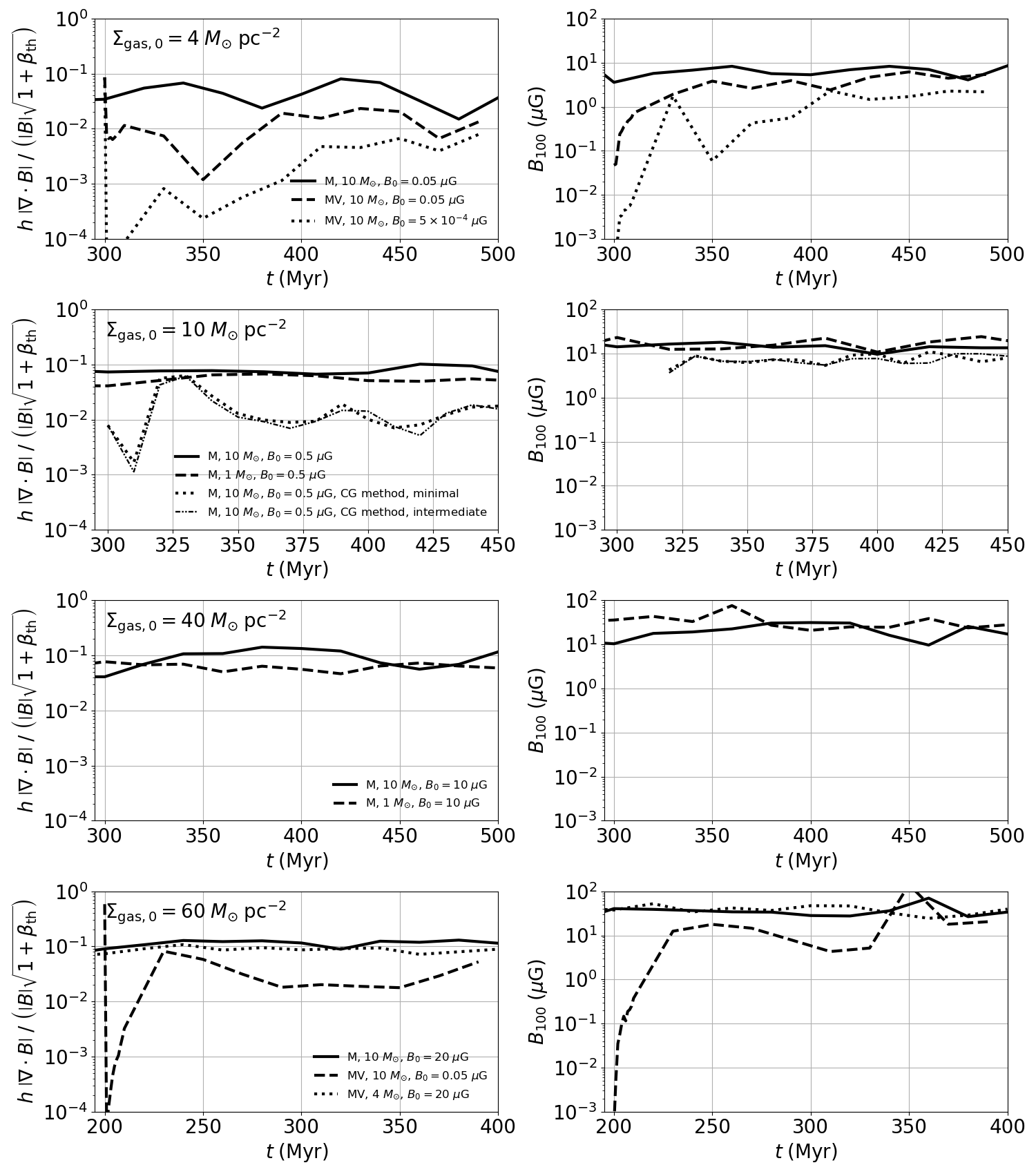}}
	\caption{
Left column: relative errors in $\nabla \cdot \bs{B}$ as a function of time for a subset of our simulations. The initial gas surface density is indicated in the top left corner of each panel, with the value of $B_0$ and resolution indicated in the legend. Right column: as for the left column, but for $B_{100}$, the mass weighted magnetic field strength in $n>100\;\rm{cm}^{-3}$ gas.
		}
		\label{fig: divB test}
\end{figure}

In this appendix we briefly discuss the 
maintenance  of $\nabla \cdot \bf{B} \approx 0$ in our models. The left column of Fig. \ref{fig: divB test} shows the time evolution of the relative error in $\nabla \cdot \bs{B}$, while the right column shows the time evolution of $B_{100}$. 
We display a subset of models presented throughout this work, and add two variants on our M10 model at a resolution of $m_{\rm g}=10\;M_{\odot}$ in which we implement the CG method presented in \citet{Hopkins2016b}. This method iteratively approximates a globally divergence-free reconstruction of the fluid, at a modest CPU cost of $\sim20-30\%$ of the hydrodynamics computation. When used in conjunction with the \citet{Dedner2002} and \citet{Powell1999} cleaning methods, the CG method has been shown to reduce the mean error in $\nabla \cdot \bs{B}$ by $\sim1$ dex when compared to using the same cleaning methods without the CG method. We run two test simulations, one with the minimal setting, and one with the intermediate setting. These settings differ in how aggressive they are at reducing the relative error in $\nabla \cdot \bs{B}$ and in how computationally expensive they are. In order to measure the error in $\nabla \cdot \bs{B}$ we compute the average of $h\;\left|\nabla \cdot \bs{B} \right|/\left|\bs{B}\right|\sqrt{1+\beta_{\rm{th}}}$, where $\beta_{\rm{th}}$ is the ratio between thermal and magnetic pressure, and $h$ is the cell size. This error definition allows us to estimate the deviation from $\nabla \cdot \bs{B}=0$ in the regions where the magnetic field is dynamically important, as the error contribution from cells with $\beta_{\rm{th}}\gg 1$ is suppressed. We find that our models generally display errors $\sim0.1$, consistent with simulated galaxies in \citet{Hopkins2016}. In addition, we find that the CG method reduces the error by 0.5-1 dex, with no discernible difference between the two different settings. We do find, however, that $B_{100}$ saturates at a value lower by a factor of $\sim2$ in the CG variants. We also find that for our MV models, the error in $\nabla \cdot \bs{B}$, which initially has a high value due to the construction of initial conditions, is reduced to a value comparable with our M models within 1 Myr. This allows us to trust the amplification of the magnetic field in our MV models to the extent that we trust our M models, even with the $\nabla \cdot \bs{B}=0$ violating initial conditions.

\end{document}